\documentclass[11pt,a4paper]{elsarticle}

\usepackage{amssymb}
\usepackage{amsmath}
\journal{ArXiV}

\begin{document}

\begin{frontmatter}



%\title{On hyperbolicity for nerve pulse propagation in axons}
\title{The modelling of the action potentials in myelinated nerve fibres}


\author[label1]{Kert Tamm\corref{cor1}} %% Author name
\author[label1]{Tanel Peets}
\author[label1,label2]{J\"uri Engelbrecht}

\affiliation[label1]{organization={Department of Cybernetics, Tallinn University of Technology},%Department and Organization
            addressline={Ehitajate tee 5}, 
            city={Tallinn},
            postcode={19086}, 
            state={Harjumaa},
            country={Estonia}}
\affiliation[label2]{organization={Estonian Academy of Sciences},%Department and Organization
            addressline={Kohtu 6}, 
            city={Tallinn},
            postcode={10130}, 
            state={Harjumaa},
            country={Estonia}}
\cortext[cor1]{kert.tamm@taltech.ee}

%% Abstract
\begin{abstract}
The initial version of the planned paper has gone through a major revision in 2025. First, the paper ended up growing a bit too long, and as a result of that, we decided to split it into two parts. The first part focuses on the model for the unmyelinated case and its behaviour, and the second part focuses on including the influence of myelination into the model. Second, when the initial version of the manuscript was going through the review process, it became evident that the way the content was presented was somewhat confusing for readers with a background in the experimental side of research into nerve processes. As a result, we went through a major revision, redoing all the numerical simulations with parameters that are closer to what Hodgkin and Huxley used in their classical paper from 1952, where the Hodgkin-Huxley model was initially introduced. The second major change was to change the logic how the specific inductance value is chosen for the numerical example - in the previous version it was chosen by aiming for a specific propagation velocity when the axon radius was chosen as 1 micrometre, in the updated version the value is chosen to get the AP propagation velocity which was experimentally observed in the HH 1952 paper at the same parameters as were used in that paper.\\ 
\textbf{Part 1 - On hyperbolicity for nerve pulse propagation in axons.}
The classical Hodgkin-Huxley (HH) model describes the propagation of an axon potential (AP) in unmyelinated axons. The hypothesis is that the AP propagation in axons depends not only on the HH ion mechanism but also on capacitance and inductance.
In this paper, we revisit a model proposed by Lieberstein for describing propagating AP in unmyelinated axon, including the possible effect of inductance that might influence velocity, into the governing equation. In many cases the axons have a myelin sheath and the experimental studies have then revealed significant changes in the velocity of APs. Next, the goal is to modify Lieberstein model further to include the influence of myelination on the AP dynamical behaviour. However, before we can do that we have to check that the solutions of the governing equations fulfil all the essential requirements for describing the nerve signalling in a physiologically plausible way. 
The numerical simulation using the physical variables demonstrates the changes in the velocity of an AP as well as the changes in its profile. These results match well the known effects from experimental studies.\\
\textbf{Part 2 - The modelling of the action potentials in myelinated nerve fibres.}
The classical Hodgkin-Huxley model describes the propagation of an axon potential (AP) in unmyelinated axons. In many cases the axons have a myelin sheath and the experimental studies have then revealed significant changes in the velocity of APs. In this paper, a theoretical model is proposed describing the AP propagation in myelinated axons. As far as the velocity of an AP is affected, the basis of the model is taken after Lieberstein, who included the possible effect of inductance that might influence velocity, into the governing equation. The proposed model includes the structural properties of the myelin sheath: the $\mu$-ratio (the ratio of the length of the myelin sheath and the Ranvier node) and g-ratio (the ratio of the inner-to-outer diameter of a myelinated axon) through parameter $\gamma$. The Lieberstein model can describe all the essential effects characteristic to the formation and propagation of an AP in an unmyelinated axon. Then a phenomenological model (a wave-type equation) for a myelinated axon is described including the influence of the structural properties of the myelin sheath and the radius of an axon. The numerical simulation using the physical variables demonstrates the changes in the velocity of an AP. These results match well the known effects from experimental studies.
\end{abstract}


\end{frontmatter}



\newpage{\large \noindent \textbf{Part 1 -- On hyperbolicity for nerve pulse propagation in axons}}
\section{Introduction}\label{sec1p1}

The celebrated Hodgkin-Huxley (HH) model describes the dynamics of an action potential (AP) in unmyelinated axons \cite{Hodgkin1952}. This model explicitly describes the role of ion currents in forming an asymmetric AP in an unmyelinated nerve fibre.  In deriving the governing equations for a squid giant axon, Hodgkin and Huxley neglected the role of inductance and their model is based on the cable equation which is a diffusion-type equation \cite{Scott1999}. Their model (in the form of an ordinary differential equation (ODE)) was useful for calculating an AP at a certain space point in time. In order to get a propagating wave, they made a few additional assumptions to get back to governing equations that are in the form of partial differential equations (PDE). It should be noted that partly the reasoning behind preferring the ODE form of HH model was the performance of computational resources back in 1950'ies. Clearly, the computational resources have increased by many orders of magnitude over the past 75 years. Alternatively, based on part of the work done by Hodgkin and Huxley \cite{Hodgkin1952}, Lieberstein \cite{Lieberstein1967} proposed a model that allows a propagating AP signal by opting to keep the inductivity in a way that does not require these additional assumptions. 

\subsection{Lieberstein hypothesis}
Lieberstein \cite{Lieberstein1967} hypothesis -- the AP propagation in axons depends not only on the HH ion mechanism but also on capacitance and inductance. 

\subsection{The question of inductance in the context of nerve fibres}
The discussion of the role of inductance started even before the derivation of the HH model. Cole has analysed the influence of inductance in the process of forming signals in axons \cite{Cole1941,ColeBaker1941} and concluded that it may be worth including inductance in the model. Lieberstein has modified the HH model by returning to basics and kept inductance in the model \cite{Lieberstein1967} and added the mechanism of ion currents proposed by Hodgkin and Huxley \cite{Hodgkin1952}. In this model, the governing equation is hyperbolic and describes a propagating wave. Having a hyperbolic governing equation (wave-equation type) is beneficial as it ensures that the process remains causal (i.e., velocity is finite) as opposed to parabolic (heat-equation type) where the signal can have infinite propagation velocity (i.e., breaking causality). Recently, Wang et al \cite{Wang2021} have stressed the importance of inductance in processes of myelinated axons. In this context, the Lieberstein model may be more informative describing the process in more detail. The original governing equations of Lieberstein \cite{Lieberstein1967} were solved by the finite difference method and the results obtained by numerical simulation demonstrated some wave profiles in time. Below we shall explain the derivation of the Lieberstein governing equations and demonstrate that the model can describe all the essential effects characteristic to the formation and propagation of an AP in unmyelinated axons. In addition, the profiles of ion currents and phenomenological variables proposed by Hodgkin and Huxley \cite{Hodgkin1952} are analysed based on the solution of the Lieberstein equation. In this way, the results of this study complete the analysis of the Lieberstein model preparing the ground for the modelling of the AP in the myelinated axons.

\subsection{Structure of the paper}\label{subsec3}
We start with a brief overview of the structure of nerve cells in Section \ref{subsec1} and a brief overview of classical models in such a context in Section \ref{subsec2}.
Section \ref{sec2} is devoted to the analysis of the Lieberstein  \cite{Lieberstein1967} model for the unmyelinated axon including the brief overview of assumptions made when deriving the noted model. 
Section \ref{numresults} contains the profiles of an AP found by the numerical simulation using the physical units  \cite{Hodgkin1952,Lieberstein1967} demonstrating the accuracy of the model. 
The obtained results confirm the earlier analysis by Kaplan and Trujillo \cite{Kaplan1970} but more details are given. The profiles of ion currents and changes of the phenomenological variables $n$, $m$, and $h$ \cite{Hodgkin1952} during the propagation of the AP are also presented. It is shown that the head-on-collision of two APs leads to annihilation. The influence of the refractory period for the formation of consecutive APs is also analysed as well as the dependence of the velocity on the diameter of the axon. These well-known effects demonstrate the validity of the Lieberstein model and encourage us to use this equation for the modelling of the APs. Finally, in Section \ref{discussion} the discussion is presented with conclusions.  As far as the numerical simulations are carried out with physical variables (i.e., in physiologically viable range), the results could be better checked in experimental studies.

\section{Brief description of the physical structure of an axon}\label{subsec1}

\begin{figure}[h]
\includegraphics[width=0.99\textwidth]{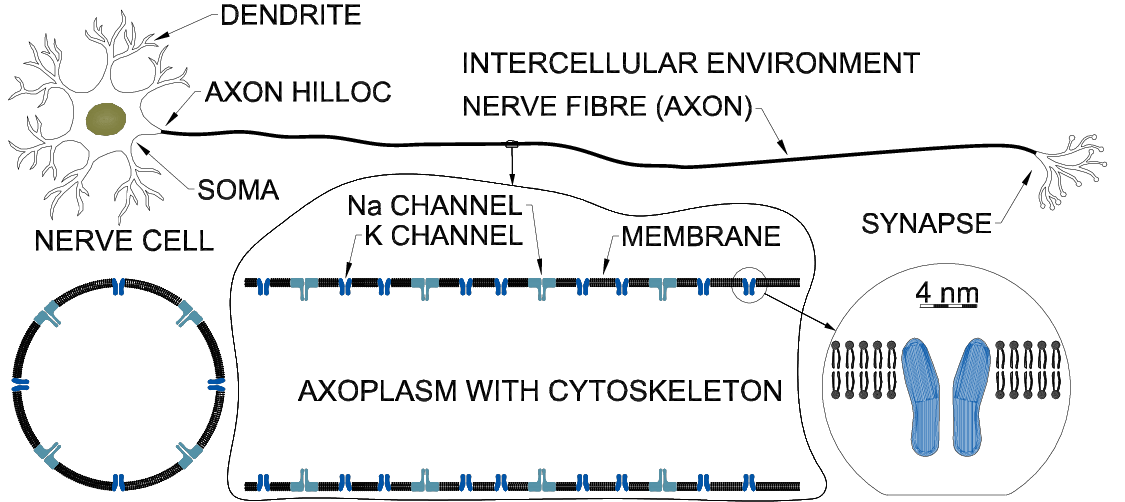}
\caption{The artistic sketch of a nerve cell and a structure of an unmyelinated axon.}\label{axonskeem}
\end{figure}
The main structural element in neural networks is the axon along which an electrical signal (the action potential (AP)) propagates from the cell body to the nerve terminal. An axon can be modelled as a tube in a certain environment \cite{Huxley1949}. Inside the tube is the axoplasmic fluid (shortly axoplasm), and the wall (in terms of continuum mechanics) of the tube has a bi-layered lipid structure called biomembrane. The biomembrane is composed of two layers of amphiphilic phospholipids and also includes proteins which are responsible for forming the ion channels. Such a simple biomembrane is called unmyelinated but in many cases, the biomembrane has an additional myelin sheath composed of multiple layers of a glial membrane. In this case, the axon is called myelinated. The sheath is interrupted by nodes of Ranvier where ion channels are active like in the case of unmyelinated axon. Under the myelin sheath, typically there are no active ion channels. The idealised schemes for a nerve cell and unmyelinated axon are shown in Fig.~\ref{axonskeem}.

\section{Modelling}\label{subsec2}
The strength of the HH model is in the detailed description of the mechanism of ion currents but the cable equation describing the propagation of an AP is simplified by neglecting the inductance. There are several ways to improve this classical model towards a better description of the structural properties of axons and taking into account the accompanying effects. Many studies are devoted to the description of mechanical and/or thermal effects accompanying the propagation of an AP  \cite{Raamat2021,ElHady2015,Chen2019,Kang2020} resulting in the formation of an ensemble of waves. These theoretical models are based on experimental results  \cite{Iwasa1980,Tasaki1988,Tasaki1989,Yang2018,Terakawa1985,Tasaki1992}. %Another important avenue of studies is related to the modelling of the behaviour of an AP in myelinated axons. 

The starting point for the modelling of an unmyelinated axon is a cylindrical tube embedded into the extracellular fluid. The barrier between extra- and intracellular fluids is a lipid bilayer with proteins. Such a biomembrane is sometimes modelled like a simple lipid bi-layer only \cite{Raamat2021}.  
In the case of the myelinated axon, this {barrier} has a myelin sheath which consists of multiple layers of a glial membrane composed of lipids and proteins and serves as an insulator  \cite{Debanne2011} or in some scenarios even as signal modulating element \cite{Tomassy2014,Fields2014a}. It means that under the myelin sheath the ion currents through the basic biomembrane (barrier) proposed by Hodgkin and Huxley  \cite{Hodgkin1952}, do not exist. However, the myelin sheath is interrupted by Ranvier nodes where the usual HH model is adequate.  It is proposed that under the myelin sheath, the passive cable equation (the diffusion-type equation) describes the process and in Ranvier nodes, the classical HH model can be used  \cite{FitzSalt,Goldman1968}. However, it is demonstrated experimentally that neurons which have myelin sheathing often show higher AP velocities compared to similar unmyelinated neurons.
This effect was already reported by Lillie in 1924 \cite{Lillie1925} and is nowadays referred to as the Lillie transition  \cite{Young2013} or saltatory conduction (see \cite{Huxley1949}). Before including the effect of myelination on the AP propagation we have to map the behaviour of the model in the simpler unmyelinated case. 

In this paper, a model based on Lieberstein \cite{Lieberstein1967} is studied which describes the propagation of an AP in an axon. This is the first step, we map the behaviour of the solutions in the unmyelinated case before we move to modify the model for the myelinated case in the future. 
The model is nonlinear and we follow a recommendation of Whitham  \cite{Whitham1974}: ``... one should not always turn too quickly to a search for the $\varepsilon$'' In the context of this model, it means that we should keep the neglected inductance as small as it is. This idea is supported by Wang et al \cite{Wang2021} who have argued that inductance is ``a missing piece of neuroscience''.  
In general terms, however, inductance means that ``there is a kind of biological structure that can store energy in a non-electrical form'' \cite{Wang2021}.
If we keep the inductance in the cable equation then the propagation is described not by a diffusion (see  \cite{Goldman1968}) equation but by a hyperbolic equation. This hypothesis may better explain the changes in the velocity of an AP. 
Noting the dependence of velocity on the diameter of a fibre \cite{Rushton1951} in the earlier studies and this relationship should also be taken into account.

In what follows, the model of Lieberstein \cite{Lieberstein1967} is a basic one to describe the AP propagation in axons. This model includes the HH ion currents but in addition, includes the inductance as it follows from the Maxwell equations \cite{Lucht2014}. 
%The main idea of this study is based on the following considerations. When dealing with a nonlinear system, we follow Whitham's idea \cite{Whitham1974} that all possible small terms (influences) should be taken into account. Moreover, in constructing a mathematical model for the propagation of an AP in axons, it is better to start from the first principles of physics perspective, i.e., from Maxwell equations \cite{Lucht2014}. 
However, before moving on to the cases with higher complexity (i.e., including effects of myelination in the future) we should make sure that the basic model captures all the essential effects of the AP propagation. 
%And, lastly, if the model requires improvements where the current knowledge does not allow precise theoretical formulation, it is acceptable to use phenomenology as a good enough place-holder, in essence, until the responsible processes can be explicitly modelled. The general principles of phenomenological modelling are described by Engelbrecht et al \cite{Engelbrecht2024}.

\section{Lieberstein model for AP propagation on unmyelinated axon}\label{sec2}

The goal is to model propagation of AP along axon which is a component of the nerve cell located between the cell body and synapse (where the signal is transmitted to the next nerve cell, see also Fig.~\ref{axonskeem}). 
%As  outlined above we follow here the work of Lieberstein, who started similarly to Hodgkin and Huxley from Maxwell equations but opted to keep the inductivity and as such arrives at a model in the PDE (hyperbolic) form, unlike classical HH model which is usually presented in the ODE form and would need some additional assumptions to get back into hyperbolic form. 
Lieberstein model can be considered as a modification of the HH model as underlying major assumptions during the derivation are the same, like, for example, description of the dynamics of key protein channels (ion channels). 
%As noted before, when deriving a mathematical model, all possible phenomena that could influence the propagation velocity should be taken into account, even if some of these could be considered much smaller than others at first glance (see, for example, Whitham \cite{Whitham1974}). 
Here we shall demonstrate that the Lieberstein model is able to describe all the important phenomena in unmyelinated axons and will be a great canditate for further modifications for the purpose of modelling AP propagation in myelinated axons.

\subsection{Basic model}
The cable equation \cite{FitzSalt} used by Hodgkin and Huxley \cite{Hodgkin1952} for deriving the {equations} for the electrical signal propagation is parabolic, i.e., inductance is neglected. Indeed, its influence is small but following Whitham \cite{Whitham1974} we use here the full model of Lieberstein \cite{Lieberstein1967} derived directly from Maxwell equations \cite{Lucht2014}. This model is hyperbolic but the final velocity of the signal is influenced by ion currents like in the HH {equations} \cite{Hodgkin1952} as noted before. 
 
Here we go briefly over the key points from \cite{Lieberstein1967} 
demonstrating the importance of a threshold and annihilation and presenting the profiles of ion currents and phenomenological variables $n$, $m$, $h$ (not presented in \cite{Lieberstein1967}).
We add some comments related to the modifications we plan to do in the future when including the effect of myelination on the AP propagation along the axon.

Lieberstein \cite{Lieberstein1967} starts with an elementary form of Maxwell equations for current and voltage on a long line
\begin{equation}\label{lib1}
\frac{\partial i_a}{\partial x} + i + \pi a^2 C_a \frac{\partial V}{\partial t} =0,
\end{equation}
\begin{equation}\label{lib2}
\frac{\partial V}{\partial x} + r i_a + \frac{L}{\pi a^2}\frac{\partial i_a}{\partial t} =0,
\end{equation}
where 
\begin{itemize}
\item $x$ is space (length) and $t$ is time;
\item $V$ is the action potential, $i_a$ is the line axon current (along the axon) and $i$ is the membrane current per unit length (taken the same as HH current across the membrane later in the paper);
\item $a$ is the radius of the axon, $r$ is the axon resistance per unit length, $L$ is the axon specific self-inductance and $C_{a}$ is the axon self capacitance per unit area per unit length.
\end{itemize}
It is noted that the membrane current density (which is what is used in the HH model \cite{Hodgkin1952}) is
\begin{equation}
I = \frac{i}{2 \pi a},
\end{equation}
and the specific resistance of the axon is
\begin{equation}
R = \pi a^2 r.
\end{equation}

It is easier to understand the structure and the physical interpretation of governing equations \eqref{lib1} and \eqref{lib2} by transferring the system of equations to a single second-order PDE (see eq.~(3) in \cite{Lieberstein1967}):
%The structure and the physical interpretation of the governing equations \eqref{lib1} and \eqref{lib2} is easier to see if writing it as a single second-order PDE (see eq.~(3) in \cite{Lieberstein1967}):
\begin{equation}
\frac{\partial^2 V}{\partial x^2} - L C_a \frac{\partial^2 V}{\partial t^2} = R C_a \frac{\partial V}{\partial t} + \frac{2}{a} R I + \frac{2}{a} L \frac{\partial I}{\partial t},
\end{equation}
where on the left-hand side (LHS) is a classical wave equation type operator while on the right-hand side (RHS) are dissipation-type operators (first order partial time derivatives). These operators either remove energy from the system or add it to the system depending on the sign of the operator. In the context of signal propagation in nerve fibres, the ionic currents across the membrane change the membrane potential in a given location while some of that energy is lost during the diffusion-type propagation along the axon to the axon resistance. At the same time, the LHS operator ensures that part of the signal is propagating like a classical wave.

Returning to the system as two coupled eqs.~\eqref{lib1} and \eqref{lib2}, the membrane current $I$ is taken (this is what is different from regular telegraph equation) as it was proposed in the HH model \cite{Hodgkin1952} as
\begin{equation} \label{HHcurrent}
I = C_m \frac{\partial V}{\partial t} + I_{Na} + I_{K} + I_l,
\end{equation}
where $I$ is the total membrane current density (positive direction is into the axon), $V$ is the membrane potential relative to the resting potential (depolarisation is negative), $C_m$ is the membrane capacity per unit area while $I_{Na}$ and $I_K$ are currents which are defined through ``equilibrium potentials'' for Na and K ions while $I_l$ is ``leakage current'' representing \emph{all} other ionic currents (Hodgkin and Huxley refer to chlorine Cl, however, Ca is also relevant for some excitable cells) and is chosen so that at resting potential leakage current across the membrane would be zero. 

In eq.~\eqref{lib1} we have membrane current per unit length $i$, which is related to the membrane current density as $i = 2 \pi a I$ and if one takes $I$ as expressed in eq.~\eqref{HHcurrent} inserting it into eq.~\eqref{lib1}, we can get expression
\begin{equation}\label{libvahe}
C_a \pi a^2 \frac{\partial V}{\partial t} + \frac{\partial i_a}{\partial x} + 2 \pi a \left( C_m \frac{\partial V}{\partial t} + I_{Na} + I_{K} + I_l \right)  = 0.
\end{equation}
Further, inserting $I_{Na}, I_{K}, I_l$ \cite{Hodgkin1952} and collecting both time derivatives in eq.~\eqref{libvahe}, we get:
\begin{equation} \label{LIB11}
\begin{split}
& \left(C_a \pi a^2 + C_m 2 \pi a\right)\frac{\partial V}{\partial t} + \frac{\partial i_a}{\partial x} +\\
&  + 2 \pi a \cdot \left[\hat{g}_K n^4 (V-V_K) + \hat{g}_{Na} m^3 h (V-V_{Na}) + \hat{g}_l (V-V_l)  \right] = 0,
\end{split}
\end{equation}
and from eq.~\eqref{lib2}:
\begin{equation} \label{LIB21}
\frac{L}{\pi a^2}\frac{\partial i_a}{\partial t} + \frac{\partial V}{\partial x} + r i_a = 0.
\end{equation}
Equations~\eqref{LIB11} and \eqref{LIB21} can be solved numerically with the pseudospectral method. It should be noted that Lieberstein \cite{Lieberstein1967} remarks that as $a$ is small then  $C_a \pi a^2 << C_m 2 \pi a$, and the influence of $C_a$ is typically assumed to be negligible. We take $C_a = 0$ in the following numerical example. We performed also a few numerical simulations with $C_a = 0.1$ (just taking it one order of magnitude smaller than $C_m$ to check if it does anything significant if taken large enough) but observed its influence to be small on the behaviour of the numerical solutions. 

%The dimensions of coefficients in  eqs.~\eqref{LIB11} and \eqref{LIB21} are: $C_a \left[\frac{\mathrm{F}}{\mathrm{m^3}}\right]$, $C_m \left[\frac{\mathrm{F}}{\mathrm{m^2}}\right]$, $a [\mathrm{m}]$, $n,m,h$ are dimensionless, $\hat{g}_K,\hat{g}_{Na},\hat{g}_l \left[\frac{\mathrm{S}}{\mathrm{m^2}}\right]$ ($\mathrm{S}$ denotes SI-unit: [Siemens]) while the membrane potential $V [\mathrm{V}]$ and {the line axon} current $i_a [\mathrm{A}]$. %Equation~\eqref{LIB1} is written in $\left[\frac{V}{s}\right]$ and eq.~\eqref{LIB21} -- in $\left[\frac{A}{s}\right]$. {As usual, the space $x$ dimension is $[\mathrm{m}]$, time $t$ dimension is $[\mathrm{s}]$, while in addition $L$ dimension is $[\mathrm{H \cdot m}]$ and $r$ dimension is $[\mathrm{\Omega}/ \mathrm{m}]$.}
%{The space} $x [\mathrm{m}]$, {time} $t [\mathrm{s}]$ {and} $L [\mathrm{H} \cdot \mathrm{m}]$ {while} $r .

The difference in keeping the inductance $L$ contrary to the celebrated classical HH model \cite{Hodgkin1952} is the following. One possible way of making sense of the differences is by looking at the equations. In the case of the system of PDEs (eqs.~(1) and (2) in \cite{Lieberstein1967}), it can be seen that there is a term $\partial i_a / \partial t $ in eq.~(2). This can be interpreted as the rate of change of current along an axon. In the HH {equations}, the axon current is ``hidden'' in term $\partial^2 V / \partial x^2$. Another possibility is to look at the Lieberstein model in the form of one PDE (eqs.~(3) and (6) in \cite{Lieberstein1967}). It can be seen in eq.~(3) that two additional terms appear in the Lieberstein model compared to the classical HH {equations} -- $\partial^2 V / \partial t^2 $ and $\partial I / \partial t$. The second partial derivative is what makes the model hyperbolic and as pointed out by Kaplan and Trujillo \cite{Kaplan1970}, it influences the maximum velocity of the AP. While the term $\partial I / \partial t$ is another notable difference as the time derivative of the ion current is not present in the classical HH model. Observing eq.~(6) in \cite{Kaplan1970} it can be seen that inductance affects the ion currents and also time derivatives of gating variables $n, m, h$ arise. One should note that Kaplan and Trujillo \cite{Kaplan1970} have calculated axon inductance from the movement of ions as $4421\,[\mathrm{mH} \cdot \mathrm{cm}]$ if AP velocity is $12.3\,[\mathrm{m/s}]$ at $a=238\,[$\textmu  $\mathrm{m}]$ {and Cole and Baker have also investigated question of inductance in this context} \cite{Cole1941,ColeBaker1941}. It can be noted that similar models involving inductivity are in use in cardiophysiology, see, for example, \cite{Rossi2017}.

\subsection{Moving frame of reference and inductivity}

Lieberstein \cite{Lieberstein1967} assumes that accounting for the inductivity (unlike in the HH {equations} where it was neglected), he can go into a moving frame of reference in a standard way introducing
$
\xi = x - \Theta t.
$
However, we should note that it could be, actually $\xi = x \pm \Theta t$, where $\Theta$ is velocity, as taking only a single direction means discarding the wave propagating in the opposite direction. We remark that using a moving frame of reference allows one to also derive an evolution equation (a single wave) for the nerve pulse \cite{Engelbrecht1981}. Lieberstein \cite{Lieberstein1967} defined the velocity $\Theta$ of the AP in an unmyelinated axon through axon radius $a$, inductance $L$ and capacity $C$ as
\begin{equation} \label{libkiirus}
\Theta = \sqrt{\frac{a}{2 L C}},
\end{equation}
proceeding after that to writing up eqs \eqref{lib1} and \eqref{lib2} in the moving frame of reference. For the sake of completeness, it should be noted that in \eqref{libkiirus} Lieberstein \cite{Lieberstein1967} argued that for capacitance $C$ the axon self-capacitance $C_a$ contribution can be neglected because $C= \frac{a}{2} C_a + C_m$ and axon radius is small so the contribution of $C_a$ is small compared to membrane capacity per unit area $C_m$.

More recently, Fukasawa and Takizawa \cite{Fukasawa2016} have also investigated the problem of electrical signal propagation velocity in an axon if inductivity is taken into account including the question of how to estimate the inductance parameter value. 

In the present paper, we have opted to avoid going into a moving frame of reference preferring to solve the model as a coupled pair of PDEs and to keep the influence of $C_a$ for the sake of completeness, even if its influence on the behaviour of the solutions is arguably small. We comment that first, for our chosen numerical solving method it is more convenient to solve a coupled pair of PDEs instead of a single higher order PDE with mixed partial derivatives and second, the model will be easier to modify in the future to include effects of myelination in the paired PDE form where the action potential across the membrane and ionic currents along the axon are separately represented.

\section{Solutions of Lieberstein model}\label{numresults}

The system is solved using the pseudospectral method (PSM) (see Appendix A in \cite{Raamat2021} for details) and periodic boundary conditions demonstrating the evolution of solutions for model equations \eqref{LIB11} and \eqref{LIB21} (see Fig.~\ref{fig1}). For initial condition we generate a narrow localised pulse for AP in the middle of the 1D space domain at time $T_0=0$ as
%\begin{equation}
$V(X,T_0) = V_0 \mathrm{sech}^2 (B_0 \cdot X_0),$ where $X_0 = X - l_0 \cdot \pi,$
%\end{equation}
where $V_0$ is amplitude of the pulse (-15 [mV] unless stated otherwise), $B_0 = 0.5$ is the width parameter of the pulse, $l_0 = 12$ is number of 2$\pi$ sections in space. For axial current we take zero initial value and values for $n,m,h$ are given in Table 1. 
Briefly, the main point of the pseudospectral method is that the discrete Fourier transform (DFT) based (PSM) (see also 
\cite{Fornberg1998}) can be used to represent variable $V$  in the Fourier space as
\begin{equation} \label{dft11}
\widehat{V}(k,T) = \mathrm{F} \left[ V \right]= \sum^{n-1}_{j=0}{V(j \Delta X, T) \exp{\left(-\frac{2 \pi \mathrm{i} j k}{n} \right)}},
\end{equation}
{where $n$ is the number of space-grid points, $\Delta X=2 \pi/n$ is the space step, $k=0,\pm1,\pm2,\ldots,\pm(n/2-1),-n/2$; $\mathrm{i}$ is the imaginary unit, $\mathrm{F}$ denotes the DFT and $\mathrm{F}^{-1}$ denotes the inverse DFT.
The idea of the PSM is to approximate space derivatives by making use of the DFT}
\begin{equation} \label{dft21}
\frac{\partial^{m} V}{\partial X^{m}} = \mathrm{F}^{-1}\left[(\mathrm{i} k)^{m} \mathrm{F}(V) \right],
\end{equation}
reducing, therefore, the PDE to an ODE and then using standard ODE solvers for integration in time. 
For integration in time, the model equations are rewritten as a system of first-order ODEs and a standard numerical integrator is applied. In the numerical examples given in the present paper the ODEPACK FORTRAN code (see 
\cite{ODE}{) ODE solver is used through its NumPy implementation. Handling of the data and initialization of the variables is done in Python by making use of the package SciPy (see }
\cite{SciPy}{) and the numerical results are analysed and visualised in the Matlab environment.}

The following parameters are used: $n=2^{13}$ (number of spatial nodes), $C_m= 1 \, [$\textmu $\mathrm{F / cm^2}]$ (membrane capacitance), $a=1\ldots 500\, [$\textmu $\mathrm{m}]$ (axon radius), $R = 35,4\, [\mathrm{ \Omega \cdot cm}]$ (axoplasm resistance) while the HH model parameters are the same as taken in \cite{Hodgkin1952}: $\hat{g}_{Na} = 120 \,[\mathrm{m.mho / cm^2}],$ $\hat{g}_K = 36 \,[\mathrm{m.mho / cm^2}],$ $\hat{g}_l = 0.3 \,[\mathrm{m.mho / cm^2}]$ and $V_{Na} = -115 \,[\mathrm{mV}],$ $V_K = 12 \,[\mathrm{mV}],$ $V_l = -10.613 \,[\mathrm{mV}]$ while $h_0 = 0.596,\, n_0 = 0.318,\, m_0 = 0.052$ (initial values for parameters $h,\, n,\, m$ in HH {equations} at $t=0$). In the following example we take the axoplasm capacitance as $C_a = 0.0 \, [$\textmu $\mathrm{F / cm^3}]$ and $r=R/(\pi a^2)$ (resistance of an axon per unit length). 
\subsection{Estimating the value of parameter $L$}
We take $L=22.2 \,[\mathrm{mH \cdot cm}]$ (inductivity) for the numerical example (resulting in 21.2 $[\mathrm{m/s}]$ propagation velocity for the AP signal with the same parameters as in the \cite{Hodgkin1952} which was experimentally observed value in that paper). 
For calculating the velocity of the AP from the numerical simulation we take a simple $\Delta x / \Delta t$ between maximum of the half-space at 5 [ms] and 10 [ms] (see Fig.~\ref{fig1}) for profiles that propagate at less than 25 [m/s] and at 3 [ms] and 5 [ms] for profiles propagating faster than 25 [m/s]. Figure \ref{fig1} is with parameters from HH classical paper \cite{Hodgkin1952} demonstrating the case where choosing $L=22.2$ [mH$\cdot$cm] yields the velocity for the AP $c_{AP}=21.2$ [m/s] which was, as noted, the experimentally measured value. It should be noted that the classical HH model was giving AP velocity of 18.8 [m/s].
We remark that, as noted earlier by Kaplan and Trujillo \cite{Kaplan1970}, have calculated axon inductance from the movement of ions as $4421\,[\mathrm{mH \cdot cm}]$ if AP velocity is $12.3\,[\mathrm{m/s}]$ at $a=238\,[$\textmu $\mathrm{m}]$ but later in the same paper they also estimate a much smaller value for $L$ based on different physical considerations. 

It can be noted that the chosen value of inductance is large {(22.2 [mH$\cdot$cm])} compared to, for example, a solid copper wire which would have self inductance of roughly 9 $\mathrm{nH}$ if the radius of the wire is 238 \textmu $\mathrm{m}$ and the length is 1 $\mathrm{cm}$. Although, clearly, the structure of the axon is very different from a solid metal fibre, the difference of roughly 6 orders of magnitude is significant. What inductivity means, in principle, can be described as a process that resists the change of current and considering various geometrical structures (like cytoskeleton) and different physical processes interacting with each other the chosen value does not appear to be outright implausible. One can note, for example, the significant mass difference between electrons and considered ions. However, it is clear that the nature of inductivity in the context of nerve fibres needs further clarification and study. We have a process in the proposed model that behaves like inductivity, {and} has a dimension related to inductivity so we have to make an assumption that it is inductivity to avoid violating Occam's razor principle -- although the relatively large value of the parameter needed to explain the experimental observations hints that there might be something more going on (see also Wang et al.~\cite{Wang2021}).
\subsection{Numerical example}
Initially we generate a narrow bell-shaped pulse (``spark'') for the $V$ in the middle of the space domain which generates the propagating AP. In spatial units, the length of the computation node ($\Delta x$) is 92 $[\mu m]$ while the width of the computational domain in the space is 24$\pi [\mathrm{cm}] \approx 75.4 [\mathrm{cm}]$. {For the sake of readability, the additional equations and values of physical quantities used in numerical simulation are collected in Table 1 following} \cite{Hodgkin1952} {and} \cite{Lieberstein1967}:
%Parameter values collected from} \cite{Hodgkin1952} {and } \cite{Lieberstein1967} { or estimated by authors can be found in Table 1:}
%
%\begin{table}
\\
\begin{center}
\begin{tabular}{ | c | c | c | }
\hline
\multicolumn{3}{|c|}{Table 1: Parameters for numerical example (collected from \cite{Hodgkin1952} {and } \cite{Lieberstein1967}).} \\
\hline
 $\alpha_n = 0.01 \frac{V+10}{\exp(\frac{V+10}{10})-1} $ & $\beta_n = 0.125 \exp(\frac{V}{80}) $ & $\frac{\mathrm{d} n}{\mathrm{d} t} = \alpha_n (1-n) - \beta_n n $ \\ 
\hline
 $\alpha_m = 0.1 \frac{V+25}{\exp(\frac{V+25}{10}) -1} $ & $\beta_m = 4 \exp(\frac{V}{18}) $ & $\frac{\mathrm{d} m}{\mathrm{d} t} = \alpha_m (1-m) - \beta_m m $ \\ 
\hline 
 $\alpha_h = 0.07 \exp(\frac{V}{20}) $ & $\beta_h = \frac{1}{\exp(\frac{V+30}{10})+1} $ & $\frac{\mathrm{d} h}{\mathrm{d} t} = \alpha_h (1-h) - \beta_h h $ \\
\hline
$h_0 = 0.596 $ & $ n_0 = 0.318 $ & $ m_0 = 0.052 $ \\
\hline
$C_a = 0.0 \left[\frac{\mu \mathrm{F}}{\mathrm{cm}^3}\right]$ & $C_m = 1 \left[\frac{\mu \mathrm{F}}{\mathrm{cm}^2}\right]$ & $R = 35.4 \left[\Omega \cdot \mathrm{cm}\right]$ \\
\hline
$\hat{g}_K = 36 \left[\frac{\mathrm{m.mho}}{\mathrm{cm}^2}\right]$ & $\hat{g}_{Na} = 120 \left[\frac{\mathrm{m.mho}}{\mathrm{cm}^2}\right]$ & $\hat{g}_l = 0.3 \left[\frac{\mathrm{m.mho}}{\mathrm{cm}^2}\right]$\\
\hline
$V_K = 12 \left[\mathrm{mV}\right]$ & $V_{Na} = -115 \left[\mathrm{mV}\right]$ & $V_l=-10.613 \left[\mathrm{mV}\right]$ \\
\hline
%\multicolumn{3}{|c|}{Parameter values from \cite{Hodgkin1952} and  \cite{Lieberstein1967} or estimated by authors.} \\
%\hline
\end{tabular}
\end{center}
%\end{table}
$\quad$\\
It should be stressed, that the chosen parameters do not represent any particular model nerve as these have been collected from studies describing different experimental setups or even taken as a rough estimate (like specific inductivity $L$). Most of the parameters are taken from \cite{Hodgkin1952} corresponding to squid giant axon at about $6.3 [\mathrm{C^\text{o}}]$ while inductivity $L$ is chosen so that the AP signal would have propagation velocity of around 21.1 $[\mathrm{m/s}]$ if the axon radius $a$ is 238 [\textmu$m]$.  
\begin{figure}[h]
\includegraphics[width=0.9\textwidth]{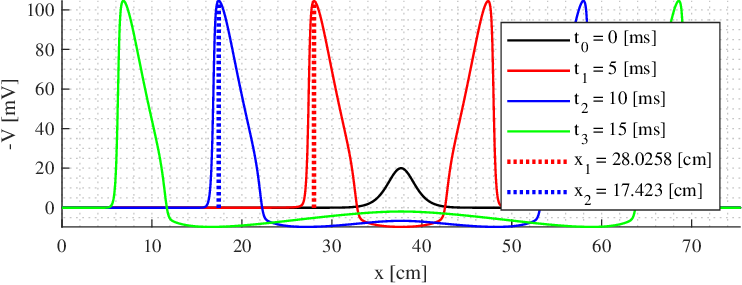}
\includegraphics[width=0.3\textwidth]{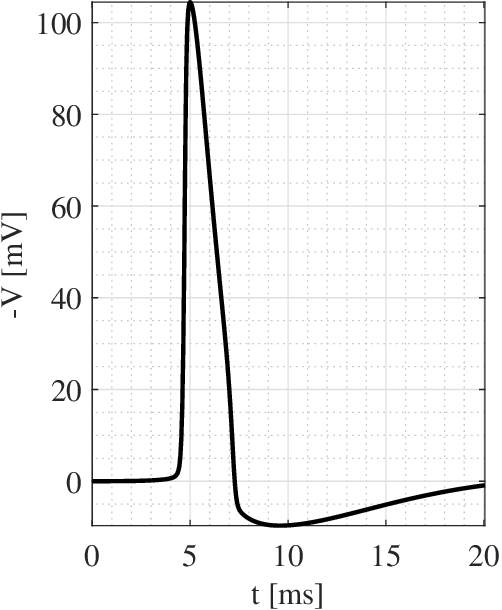}
\includegraphics[width=0.3\textwidth]{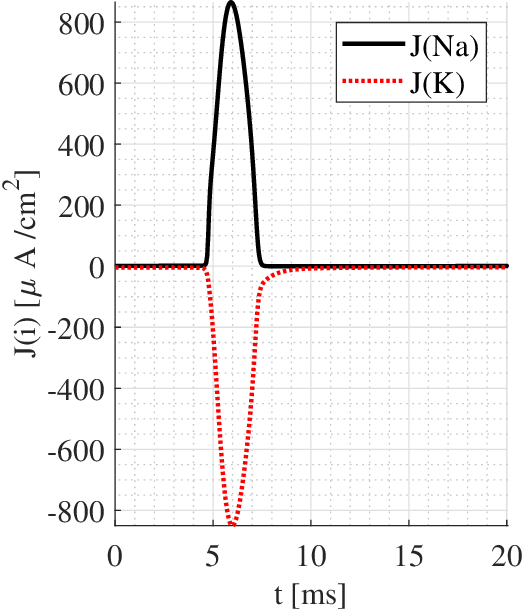}
\includegraphics[width=0.3\textwidth]{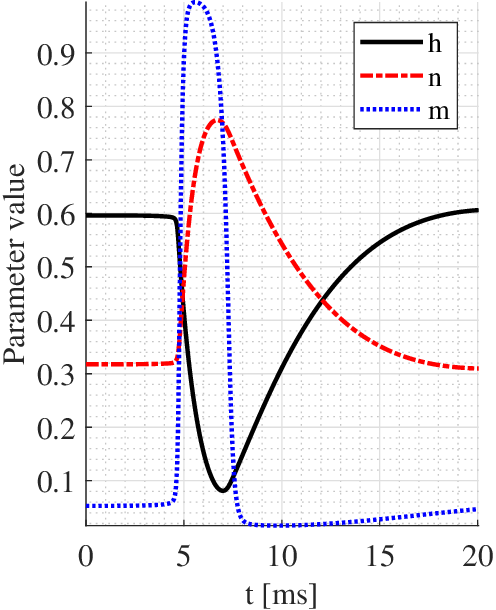}
\caption{The top panel shows AP profiles in space at 5 [ms] intervals. %Bottom panel is showing profiles in time at node $n=700$. 
The bottom panels demonstrate the example solutions of eqs \eqref{LIB11} and \eqref{LIB21} in time at $x=28$ [cm] (the $t_1=5$ [ms] location at upper panel). At the bottom the left panel shows the AP, the middle panel shows Na and K ionic currents and the right panel shows the changes of internal variables $n,\, m,\, h$ in time. For calculating velocity of the AP we take $\Delta x / \Delta t$ between maximum of the half-space at 5 [ms] and 10 [ms] for profiles that propagate at less than 25 [m/s] and at 3 [ms] and 5 [ms] for profiles propagating faster than 25 [m/s].}
\label{fig1}
\end{figure}
\begin{figure}[h!]
\includegraphics[width=0.49\textwidth]{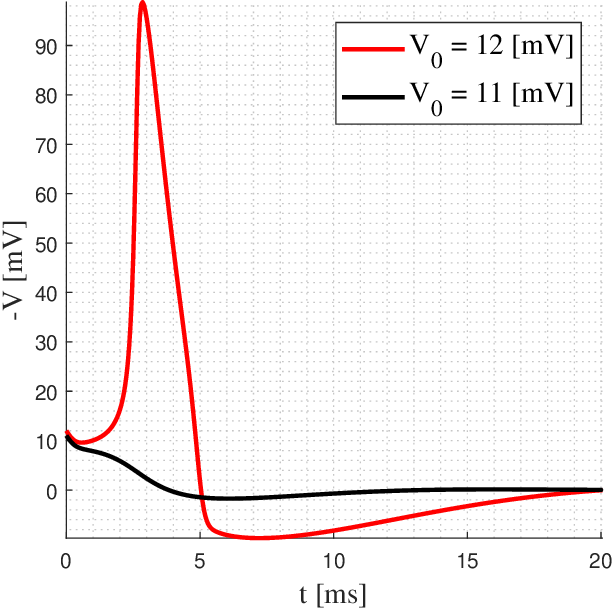}
\includegraphics[width=0.49\textwidth]{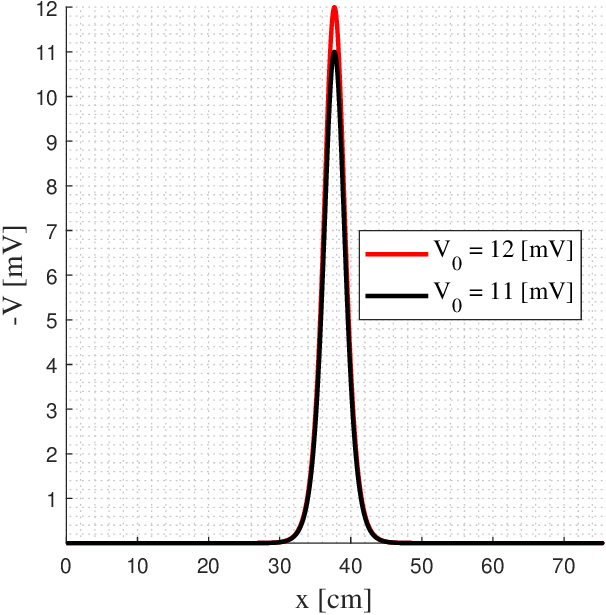}
\caption{Threshold value example. The model parameters can be found in Table 1. Left panel - AP amplitude in time at $x=37.69$ [cm] ($0.5x$ in space). Right panel - initial condition profiles with below threshold (11 [mV]) and above threshold (12 [mV] amplitudes at $t=0$ [ms]. }
\label{figTresh}
\end{figure}
\begin{figure}[h!]
\includegraphics[width=0.49\textwidth]{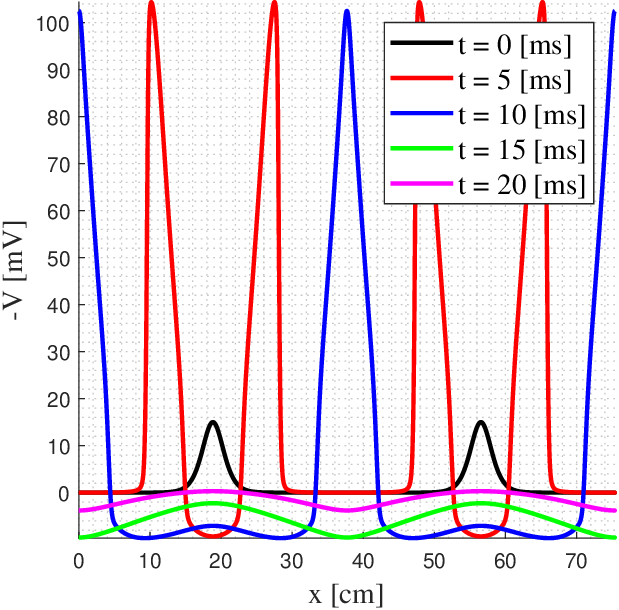}
\includegraphics[width=0.49\textwidth]{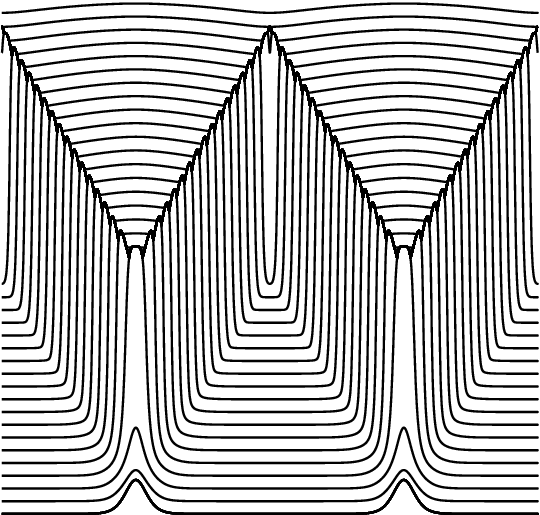}
\caption{Annihilation example of AP signals during head-on collision. The model parameters can be found in Table 1. Initial pulses given at $0.25x$ and $0.75x$ locations in space with initial amplitude of -15 [mV].}
\label{fig2}
\end{figure}
\begin{figure}[h!]
\includegraphics[width=0.49\textwidth]{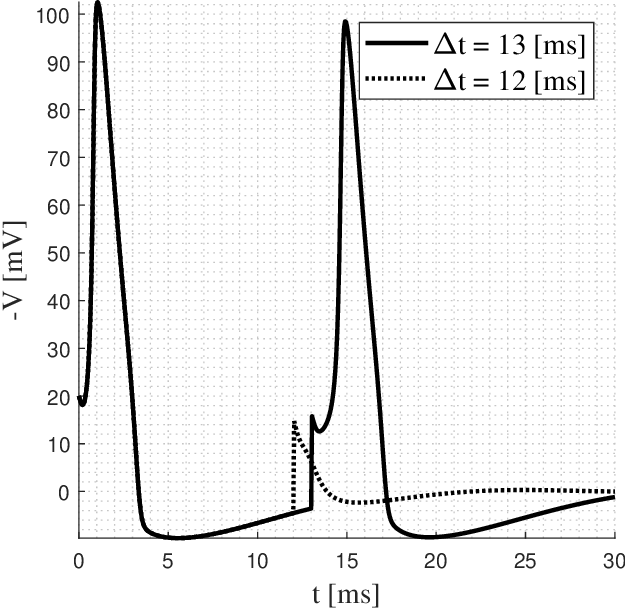}
\includegraphics[width=0.49\textwidth]{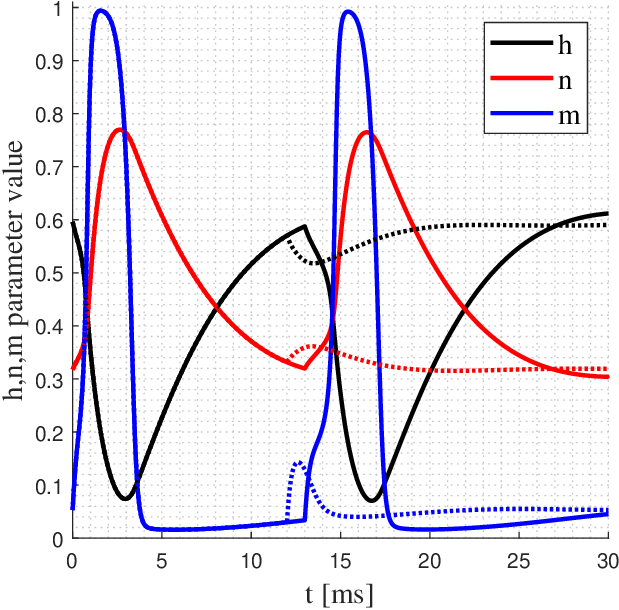}
\caption{Refraction period example. The model parameters can be found in Table 1. The pulses are given at $0.5x$ with amplitude of -20 [mV] (thershold value is roughly 12 [mV]) at 12 [ms] and 13 [ms] after the initial pulse. Left panel - the AP in time at $0.5x$ location, right panel - $n,m,h$ parameter values in time at $0.5x$ location. Solid line is pulse interval of 13 [ms] and dotted line represents pulse interval of 12 [ms].}
\label{fig3}
\end{figure}

% -----------------
%\begin{figure}[h!]
%\includegraphics[width=0.49\textwidth]{LIBHH_annihilation.eps}
%\includegraphics[width=0.49\textwidth]{LIB_timeslice_annihil.eps}
%\caption{Annihilation example of AP signals during head-on collision. The model parameters can be found in Table 1. Initial pulses given at $0.25x$ and $0.75x$ locations in space with initial amplitude of -15 [mV].}
%\label{figAnnihilation}
%\end{figure}
% -----------------
%\begin{figure}[h!]
%\includegraphics[width=0.49\textwidth]{LIBHH_refractionAP.eps}
%\includegraphics[width=0.49\textwidth]{LIBHH_refractionNMH.eps}
%\caption{Refraction period example. The model parameters can be found in Table 1. The pulses are given at $0.5x$ with amplitude of -20 [mV] (thershold value is roughly 12 [mV]) at 12 [ms] and 13 [ms] after the initial pulse. Left panel - the AP in time at $0.5x$ location, right panel - $n,m,h$ parameter values in time at $0.5x$ location. Solid line is pulse interval of 13 [ms] and dotted line represents pulse interval of 12 [ms].}
%\label{figRefraction}
%\end{figure}

In Fig.~\ref{fig1} it is demonstrated that the Lieberstein model has solutions which are characteristic of an AP in {earlier studies} (for example, \cite{Hodgkin1952,Lieberstein1967}), however, it is known that in physiology, an AP has three key properties that must be fulfilled by any valid model aiming to descibe AP. These are: (1) there exist all-or-nothing threshold above which the AP is generated and under which the signal dissipates rapidly (Fig.~\ref{figTresh}), (2) electrical nerve signal must annihilate upon head-on collision (Fig.~\ref{fig2}), (3) a disturbance or signal generated during a short time after the signal has passed a location (so-called refractory period) must dissipate rapidly without generating a new AP (a minimum separation time before another nerve pulse can propagate) (Fig.~\ref{fig3}). In Figs.~\ref{figTresh}, \ref{fig2} and \ref{fig3} it is demonstrated that the Lieberstein model satisfies these conditions. 
Figure~\ref{figTresh} demonstrates the existence of a threshold value for the Lieberstein model using the parameters mostly from \cite{Hodgkin1952} where initial excitation for the AP with amplitude of 11 [mV] dissipates rapidly while initial excitation with amplitude of 12 [mV] crosses the threshold with the used model parameters and generates the propagating AP signal. 
Figure~\ref{fig2} demonstrates that during a head-on collision the propagating AP signals annihilate according to the Lieberstein model fulfilling the second condition for a realistic AP model. The left panel of Fig.~\ref{fig2} depicts the AP profile in space with 5 [ms] intervals including the head-on collision of the profiles at 10 [ms] with the generation of two AP's at $0.25x$ and $0.75x$ while the behaviour of the signals in space-time is easier to understand from the time-slice plot on the right panel where the evolution is visualised with 0.5 [ms] intervals with horisontal axis representing space and vertical axis time. 
Figure~\ref{fig3} demonstrates existence of refraction period for the considered model where a disdurbance with 20 [mV] amplitude (threshold value was approx 12 [mV]) fails to generate a propagating AP signal when given 12 [ms] after the initial pulse but suceeds generating the propagating AP signal when given 13 [ms] after the initial pulse. These examples together with the results of Lieberstein \cite{Lieberstein1967} complete the full analysis of an improved model for an AP in unmyelinated axon. 
 
\subsection{Influence of the axon radius on the propagation velocity of the AP.}
\begin{figure}[h]
    \includegraphics[width=0.99\textwidth]{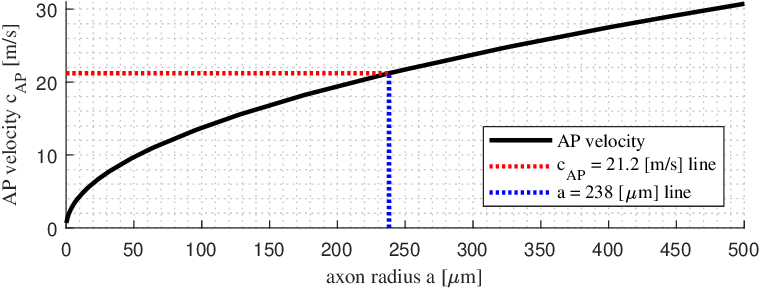}
%    \begin{tabular}{|{c}|{c}|{c}|{c}|{c}|{c}|{c}|{c}|{c}|} %{c}|{c}|{c}|{c}|{c}|{c}|{c}|
%    \hline
%    \multicolumn{9}{|c|}{Table 2: AP velocity $c_{\text{AP}}$ as a function of axon radius $a$.} \\
%      \hline
%      a $[$\textmu $\mathrm{m}]$ & 0.25 & 0.50 & 1 & 2 & 4 & 8 & 16 & 25   \\
%      \hline
%      $c_{\text{AP}}$~$[\mathrm{m/s}]$ & 0.5 & 0.75 & 1.0 & 1.4 & 1.95 & 2.55 & 3.2 & 3.55     \\
%      \hline \hline
%      a $[$\textmu $\mathrm{m}]$ & 40 & 50 & 60 & 80 & 100 & 125 & 150 & 175 \\
%      \hline
%      $c_{\text{AP}}$~$[\mathrm{m/s}]$ & 3.9 & 4.0 & 4.1 & 4.25 & 4.3 & 4.35 & 4.4 & 4.4 \\
%      \hline
%    \end{tabular}
\caption{The AP propagation velocity as a function of the axon radius in the Lieberstein model.}\label{APvel}
\end{figure}

From earlier studies (see, for example, \cite{Massey2022}) it is known that the propagation velocity of the AP is influenced by three factors -- (i) space constant $\lambda = \sqrt{R_m/R_a}$, (ii) time constant $\tau = R_m C_m$, and (iii) the time measure $T$ characterising the time it takes to generate the AP on any point along the axon. The membrane resistance $R_m$ scales inverse-proportionally with axon diameter $R_m \propto 1/d$  while membrane capacitance $C_m$ increases proportionally with axon diameter $C_m \propto d$ while axial resistance $R_a$ scales as $R_a \propto 1/d^2$ meaning that the propagation velocity of AP is $c_{AP} \propto \sqrt{d}$.

We note that in the model parameters, the axon resistance is changed when the axon radius is changing so $r=R/({\pi}a^2)$, where $R = 35.4\, [\mathrm{\Omega \cdot cm}]$ is constant \cite{Hodgkin1952}. The propagation velocity of AP as a function of axon radius is depicted in Fig.~\ref{APvel} with the parameters used for a numerical example. It should be noted that the electrical circuit hypothesis for the AP propagation velocity (like {$c_{AP}^{2}\propto K a/(2R_2 C_m)$} for the HH {equations} \cite{Hodgkin1952} or $c_{AP}^{2}\propto a/(2LC)$ for the Lieberstein model \cite{Lieberstein1967}) which is based on similarity with the chosen electrical circuit, while certainly useful, can overlook some potentially important mechanisms (like, for example, the influence of cytoskeleton) that are left aside during the many simplifications needed to get a simple electrical scheme.
 This means that in practical applications one has to consider carefully any conflicts between experimental observations and modelling results to determine if the model is sufficient for adequately describing the observed process. %For example, here we have chosen $L$ so that at $a=1$ {[\textmu$\mathrm{m}$]} the AP velocity $c_{\text{AP}}$ is $\approx 1$~$[\mathrm{m/s}]$ as a demonstration of the model behaviour, for practical applications that parameter would need to be determined from the electrophysiology of the problem under investigation as in reality, this is not a free parameter. 
It should be stressed that here the method for calculating the velocity (see Fig.~\ref{fig1}) of the AP profile is different than was used in the \cite{Hodgkin1952} or \cite{Kaplan1970} where the authors used an velocity estimate for the moving frame of reference while here we have used the propagation velocity measured from the numerical solution (including the nonlinear effects) by tracking the coordinate of the maximum of the AP profile at two different time moments and then taking $\Delta x / \Delta t$ to find the propagation velocity.

\section{Discussion and summary}\label{discussion}

Starting from the elementary form of Maxwell equations \eqref{lib1}, \eqref{lib2} and drawing inspiration from the classical HH paper \cite{Hodgkin1952} we use the total membrane current density \eqref{HHcurrent} to construct the model equations for unmyelinated axon similar to the Lieberstein model \cite{Lieberstein1967}. While Lieberstein opted to go a step further by moving into a moving frame of reference we opted to stay with the form which is closer to the Maxwell equations for a transmission line eqs.~\eqref{LIB11}, \eqref{LIB21} and used the work of Lieberstein as a source of inspiration for handling the question of inductance $L$ in the context of signal propagation along the nerve axon. Using the parameters from the earlier experimental studies we investigated briefly the solutions of the noted model for the unmyelinated axon and demonstrated that the behaviour of the solutions is in the physiologically plausible range and the key characteristics of the nervous signalling are fulfilled. These are: (i) the annihilation of AP signals during a head-on collision, (ii) the existence of activation threshold, and (iii) the refraction period after signal passing. In the parameter range considered, we {have calculated} the AP signal propagation velocity $c_{AP}$ (see Fig.~\ref{APvel}) from 0.68 $[\mathrm{m/s}]$ (at $a=0.25$ [{\textmu}m]) up to about 30.74 $[\mathrm{m/s}]$ (at $a=500$ [{\textmu}m]) for the unmyelinated axon. The key difference between the classical HH model and the Lieberstein-inspired model used here is that the mechanism for signal propagation along the axon emerges like a wave as a consequence of opting to keep the inductivity $L$. While, indeed, there exist variations of the classical HH model which support AP signal propagation where the {equations are} written in the form of PDE instead of the usual ODE form (which describes signal evolution in time at a fixed spatial point). In these equations, normally, the potential-gradient-type member responsible for propagating the signal along the axon is not as clearly defined from {the viewpoint of physics} (usually some kind of abstracted diffusion-type process is used). We remark, that this is essential for the purpose of clearer physical interpretation as we make use of the model based on the elementary form of Maxwell equations which will be modified in the further studies to include the influence of myelination on the signal propagating along the axon. 

Having a relatively simple pair of PDEs which are connected to the Maxwell equations for anything involving the movement of charges in an environment could be considered superior to investigating causal connections and making predictions than something that is not as clearly connected to the {basic physical considerations}.

To sum up, we revisited a model proposed by Lieberstein \cite{Lieberstein1967}, solved a version of the model numerically with a physiologically viable set of parameters and checked that the solutions behave as an observed AP should. The results confirm the earlier studies \cite{Lieberstein1967,Kaplan1970} with more details (like, for example, profiles of ion currents, and phenomenological variables). This forms a solid foundation for modifying the presented model for AP propagation in the future to include the effect of myelination (additional structure attached to the axon surface). The modification is in progress.

\section*{Acknowledgements}

This research was supported by the Estonian Research Council (PRG 1227). J\"uri
Engelbrecht acknowledges the support from the Estonian Academy of Sciences. 
%The authors thank Prof. Antoine Jerusalem from the University of Oxford for helpful comments during the preparation of the present paper. {The authors thank the anonymous reviewers for numerous suggestions for improvements and helpful comments.}

\section*{Author contributions}
Kert Tamm contributed by writing manuscript text and composing figures, performing numerical simulations and analysis of the results. Tanel Peets contributed analysis for comparing the modified Lieberstein equation against the classical Hodgkin-Huxley {equations}. All other work (like novel ideas presented, physical interpretation of the results, proofreading, etc) is a collaborative effort equally contributed by all authors. 

\newpage
{\large \noindent \textbf{Part 2 -- The modelling of the action potentials in myelinated nerve fibres}}

\section{Introduction}\label{sec1p2}

The celebrated Hodgkin-Huxley (HH) model can describe the action potential (AP) in an unmyelinated axon taking into account sodium and potassium ion currents  \cite{Hodgkin1952}. The strength of the HH {equations} is in the detailed description of the physics of ion currents but the cable equation describing the propagation of an AP is simplified by neglecting the inductance. There are several ways to improve this classical model towards a better description of the structural properties of axons and taking into account the accompanying effects. Many studies are devoted to the description of mechanical and/or thermal effects accompanying the propagation of an AP  \cite{Raamat2021,ElHady2015,Chen2019,Kang2020} resulting in the formation of an ensemble of waves. These theoretical models are based on experimental results  \cite{Iwasa1980,Tasaki1988,Tasaki1989,Yang2018,Terakawa1985,Tasaki1992}. Another important avenue of studies is related to modelling the behaviour of an AP in myelinated axons. 
The starting point for the modelling of an unmyelinated axon is a cylindrical tube embedded into the extracellular fluid. The barrier between extra- and intracellular fluids is a lipid bilayer with proteins. Such a biomembrane is sometimes modelled like a simple lipid bi-layer only \cite{Raamat2021}.
%An unmyelinated axon can be modelled like a cylindrical tube with a wall made of a biomembrane composed of a lipid bilayer.  
In the case of the myelinated axon, this {barrier} has a myelin sheath which consists of multiple layers of a glial membrane composed of lipids and proteins and serves as an insulator  \cite{Tomassy2014,Debanne2011} or in some scenarios even as signal modulating element \cite{Fields2014a}. It means that under the myelin sheath the ion currents through the basic biomembrane {(barrier)} proposed by Hodgkin and Huxley  \cite{Hodgkin1952}, are suppressed. However, the myelin sheath is interrupted by Ranvier nodes where the usual HH model works.  It is proposed that under the myelin sheath, the passive cable equation (the diffusion-type equation) describes the process and in Ranvier nodes, the usual HH model can be applied  \cite{FitzSalt,Goldman1968}. However, it is demonstrated experimentally that neurons which have myelin sheathing often show higher AP velocities compared to similar unmyelinated neurons. This effect was already reported by Lillie in 1924 \cite{Lillie1925} and is nowadays referred to as the Lillie transition  \cite{Young2013} or saltatory conduction (see \cite{Huxley1949}).

In this paper,  a phenomenological model is proposed for describing the propagation of an AP in a myelinated axon. The most important question is how to model the possible changes in the propagation velocity of an AP due to the myelin sheath. The process is nonlinear and we follow a recommendation of Whitham  \cite{Whitham1974}: ``... one should not always turn too quickly to a search for the $\varepsilon$.'' In the context of {our} equation, it means that we should keep the neglected inductance as small as it is. This idea is supported by Wang et al \cite{Wang2021} who have argued that inductance is ``a missing piece of neuroscience''. 
They propose that the main source of the inductance is the myelin sheath. In general terms, however, they state for the inductance that ``there is a kind of biological structure that can store energy in a non-electrical {form"} \cite{Wang2021}.
If we keep the inductance in the cable equation then, mathematically, under the myelin sheath the propagation is described not by a diffusion-type (see  \cite{Goldman1968}) equation but by a hyperbolic equation. This hypothesis may better explain the changes in the velocity of an AP. 
%Once the experimental studies have revealed the dependence of the velocity on the diameter of a fibre then such a dependence should also be taken into account.
{Given reports} \cite{Rushton1951} noting the dependence of velocity on the diameter of a fibre, this relationship should be taken into account.
 
In what follows, the model of Lieberstein \cite{Lieberstein1967} is a basic one to describe the AP propagation in axons. This model includes the HH ion currents but in addition, includes the inductance as it follows from the Maxwell equations. For proper modelling of the processes in myelinated axons, the influence of the myelin sheath must certainly be taken into account. In principle, the length of a myelinated section and its thickness are the leading structural factors in the process. Consequently, one important parameter is the ratio of lengths of a node of Ranvier $l_1$  and a myelinated section $l_2$  in the form of the myelin length ratio (or the $\mu$-ratio for short) which is proportional to $l_2/l_1$. Another important parameter is the g-ratio (see  \cite{Rushton1951}) which is the ratio of the inner-to-outer diameter of a myelinated axon. These parameters take into account the structure of a myelin sheath (c.f.~analysis by Basser  \cite{Basser2004}).

{The main idea of this study is based on the following considerations. First, when dealing with a nonlinear system, we follow Whitham's idea} \cite{Whitham1974} {that all possible small terms (influences) should be taken into account. Second, in constructing a mathematical model for the propagation of an AP in myelinated axons, we start from Maxwell equations} \cite{Lucht2014}. {Third, for the description of structured properties of the axon, we use the phenomenological approach following the ideas of Hodgkin and Huxley} \cite{Hodgkin1952} {for describing the ion currents and Bressloff} \cite{Bressloff2014} {for describing the saltatory conduction by a coupling coefficient which characterises the discrete diffusion. The general principles of phenomenological modelling are described by Engelbrecht et al} \cite{Engelbrecht2024}.

Section 2 is introducing the model for the unmyelinated axon  \cite{Lieberstein1967,Tamm2025WaMot}. Section 3 describes experimental results on myelinated axons focused on velocity changes  \cite{Lodish2004,Schmidt2019,Arancibia-Carcamo2017}, This analysis permits us to propose in Section 4 a phenomenological model describing the propagation of an AP in a myelinated axon influenced strongly by the structure of a myelin sheath ($\mu$-ratio). The structure of the governing equations follows the Lieberstein model (see Section 2) where the inductance leads to a wave-like behaviour but the final velocity of an AP depends on ion currents. The numerical simulations demonstrate clearly the changes in the velocity under the myelin sheath %{as well as the changes in the profile}
 of the propagating AP. Finally, in Section 5 the discussion is presented with conclusions.  As far as the numerical simulations are carried on with physical variables {(i.e., in physiologically viable range)}, the results could be better checked in experimental studies.
 
\section{The modelling of an unmyelinated axon}
\begin{figure}[h]
\includegraphics[width=0.9\textwidth]{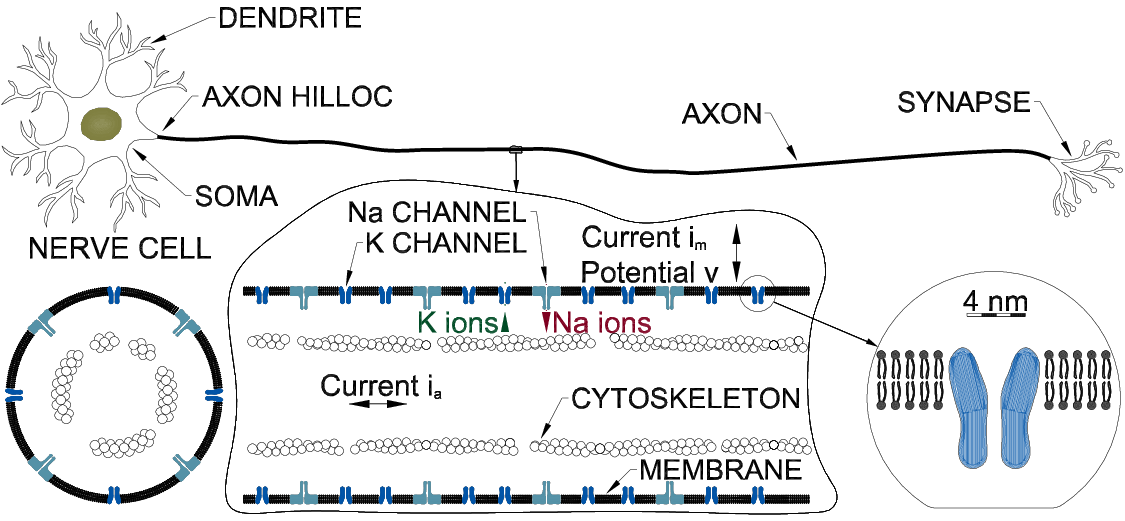}
\caption{Artistic representation of unmyelinated axon. The potential $V$ of the AP and the ionic currents $i$ act across the membrane while $i_a$ is the line axon current (along the axon). The membrane contains various proteins, like Na and K ion channels depicted here, axon is surrounded by intercellular medium and filled with axoplasm, including additional structures like cytoskeleton. The concentration of ions is different across the cell membrane. When a certain threshold is exceeded then a process is activated where Na ions flow into the axon and K ions flow out of the axon with a small time shift between the processes which is responsible for generating the AP across the membrane which typically propagates from axon hilloc towards the synapse \cite{Tamm2025WaMot,Fletcher2010,Debanne2011}. }
\label{fig_unmyl}
\end{figure}

We use a model describing AP propagation in unmyelinated axon (see Fig.~\ref{fig_unmyl}) as a starting point \cite{Tamm2025WaMot,Lieberstein1967}. First, a governing equation for describing the AP
\begin{equation} \label{LIB1}
\begin{split}
& \left(C_a \pi a^2 + C_m 2 \pi a\right)\frac{\partial v}{\partial t} + \frac{\partial i_a}{\partial x} + 2 \pi a \cdot  i_m  = 0,\\
&  i_m = \hat{g}_K n^4 (v-v_K) + \hat{g}_{Na} m^3 h (v-v_{Na}) + \hat{g}_l (v-v_l)
\end{split}
\end{equation}
and, second governing equation for describing the current along the axon
\begin{equation} \label{LIB2}
\frac{L}{\pi a^2}\frac{\partial i_a}{\partial t} + \frac{\partial v}{\partial x} + r i_a = 0.
\end{equation}
Here $x$ is space (along the axon), $t$ is time, $v$ is the action potential, $i_a$ is the line axon current (along the axon), $i_m$ is the membrane current per unit length (expressed here through dimensionless parameters  $n,m,h$, channel conductances $\hat{g}_K,\hat{g}_{Na},\hat{g}_l$ and equilibrium potentials $v_K,v_{Na},v_l$ \cite{Hodgkin1952,Lieberstein1967}), $a$ is the radius of the axon, $r$ is the axon resistance per unit length, $L$ is the axon specific self-inductance, $C_a$ is the axon self capacitance per unit area per unit length (often neglected as significantly smaller than $C_m$ but included here initially for the sake of completeness), $C_m$ is the membrane capacity per unit area. As $a$ is small then $C_a \pi a^2 << C_m 2 \pi a$ and the term  $C_a \pi a^2$ can be neglected \cite{Lieberstein1967}. 
Equations~\eqref{LIB1} and \eqref{LIB2} can be considered as a variant of the classical Hodgkin-Huxley model where the influence of specific inductance $L$ has not been neglected leading to a hyperbolic PDE (i.e., signal propagation velocity is finite) as opposed to parabolic (diffusion-type) governing equations as they arise in the HH model when inductance is dismissed. Equations~\eqref{LIB1} and \eqref{LIB2} are taken as a starting point for modification to include the influence of myelination on the AP propagation dynamics along the axon.

It should be noted that the estimated value of $L$ needed to explain the experimental observations can be relatively large, at around 4420 $mH/cm$ \cite{Lieberstein1967,MelvinLieberstein1970} as noted by Kaplan and Trujillo \cite{Kaplan1970} if AP velocity is $12.3\,[\mathrm{m/s}]$ at $a=238\,[$\textmu  $\mathrm{m}]$. Cole and Baker have also investigated question of inductance in this context \cite{Cole1941,ColeBaker1941}. It can be noted that similar models involving inductivity are in use in cardiophysiology, see, for example, \cite{Rossi2017}. Kaplan and Trujillo \cite{Kaplan1970} have also derived inductance estimate from the inertia of the ions (note that Na and K ions are several orders of magnitude more massive than electrons, as carriers of charge) finding that estimate much smaller and conclude that if there is indeed that much inductance present it can not be explained purely by the mass of the carriers of charge and there must be some other effects involved as well. We remark that inductivity is, in principle, related to a process that resists too rapid change of the current and considering various geometrical structures (like cytoskeleton) and different physical processes interacting with each other the chosen value does not appear to be outright implausible \cite{MelvinLieberstein1970}. However, it is clear that the exact nature of inductivity in the context of nerve fibres needs further clarification and study. We have a process in the proposed model that behaves like inductivity, and has a dimension related to inductivity so we have to make an assumption that it is inductivity to avoid violating Occam's razor principle -- although the relatively large value of the parameter needed to explain the experimental observations hints that there might be something more going on (see also Wang et al.~\cite{Wang2021}). We have opted to use $L=22.2 \,[\mathrm{mH \cdot cm}]$ for the numerical example (chosen to get 21.2 $[\mathrm{m/s}]$ propagation velocity for the AP signal if axon radius $a$ is 238 [\textmu$m]$) and the rest of the parameters are like they were in the Hodgkin and Huxley classical paper where the HH model was introduced \cite{Hodgkin1952}. The 21.2 $[\mathrm{m/s}]$ velocity was the experimentally observed propagation velocity and the original HH model provided velocity estimate of 18.8 $[\mathrm{m/s}]$ \cite{Hodgkin1952}.%We remark, that there are some similarities with the classical Hodgkin-Huxley model in its PDE formulation (see, for example, eq.~(29) in \cite{Hodgkin1952}) where the value of the diffusion-type coefficient determines the propagation velocity of the AP.

\begin{figure}[h]
\includegraphics[width=0.49\textwidth]{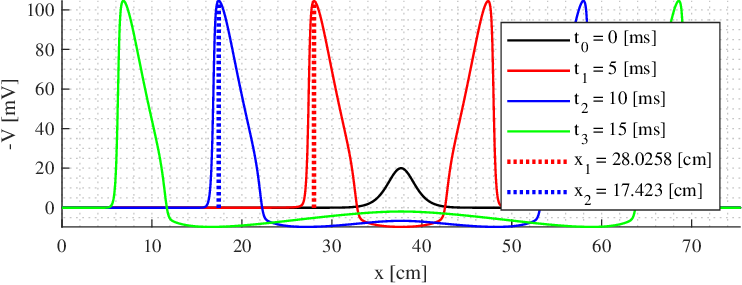}
\includegraphics[width=0.49\textwidth]{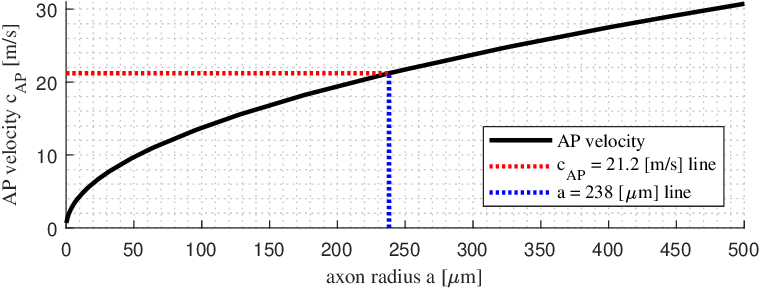}
\caption{Left panel -- for calculating the velocity of the AP from the numerical simulation we take a simple $\Delta x / \Delta t$ between maximum of the half-space at 5 [ms] and 10 [ms] for profiles that propagate at less than 25 [m/s] and at 3 [ms] and 5 [ms] for profiles propagating faster than 25 [m/s]; the parameters for the simulation are from HH classical paper \cite{Hodgkin1952} and $L=22.2$ [mH$\cdot$cm]. Right panel -- the velocity against axon radius graph for the unmyelinated case with the experimentally observed velocity and axon radius \cite{Hodgkin1952} marked by dotted lines.}
\label{velocity_unmyl}
\end{figure}

It should be stressed that here the method for calculating the velocity (see Fig.~\ref{velocity_unmyl}) of the AP profile is different than was used in the \cite{Hodgkin1952} or \cite{Kaplan1970} where the authors used an velocity estimate for the moving frame of reference while here we have used the propagation velocity measured from the solution (including the nonlinear effects) of the numerical simulation by tracking the coordinate of the maximum of the AP profile at two different time moments and then taking $\Delta x / \Delta t$ to find the propagation velocity. 

\section{Description of a myelinated axon} 
The axon can be described as a tube filled with axoplasm and cytoskeleton \cite{Debanne2011}. {The barrier between extra- and intracellular fluid has a lipid-bilayer base but also has membrane proteins} \cite{Kister2022}.
%with a lipid-bilayer (biomembrane) wall which is embedded in an intercellular medium (fluid)  
While some axons are unmyelinated, many of them have a myelin sheath (additional structure surrounding the axon formed by glial cells (oligodendrocytes (in the central nervous system) and Schwann cells (in the peripheral nervous system)) that is in simplified terms composed of layers of biomembrane glued together with some proteins.

\subsection{The structure of myelinated axon} %\label{sec3}
The myelin sheath is interrupted by the nodes of Ranvier which play an important role in the nerve pulse propagation. A node of Ranvier is typically around 1~\textmu m in length and has a high density of ion channels \cite{Lodish2004} while myelin sheath segments are typically from roughly 50~\textmu m to 300~\textmu m in length. The distribution of the myelinated parts was once believed to be uniform, but now it is understood that the length of the myelinated parts and the nodes of Ranvier vary and unusually long nodes of Ranvier (50 {\textmu}m) have been reported \cite{Tomassy2014}. These long segments could play an important role in the synchronisation of nerve pulses by delaying an AP \cite{Schmidt2019}.%{[???]}

The myelin sheath is a stack of specialised plasma membrane sheets produced by glial cells that wrap {helically} around the axon \cite{Lodish2004}. 
A lipid bilayer (biomembrane) is typically from 3 to 4 nm in thickness and {contains various proteins in and between the layers} \cite{Raasakka2019} that all together form a myelin sheath composed {of} many such layers {wrapped helically around} the axon. The myelin sheath can be up to 2.5~\textmu m in thickness \cite{Sanders1948,Michailov2004,Hanig2018} and on the lowest theoretical limit, an extra layer of biomembrane surrounding the axon might be argued to be a myelin sheath. Typical axon diameter varies from about 0.5~\textmu m to {20}~\textmu m \cite{Sanders1948} in mammals but can reach as high as about 1~mm in squid giant axon \cite{Hodgkin1949,Terakawa1985}.  For example, in peripheral nerve fibres, the myelin sheath thickness starts from about 0.5 \textmu $\mathrm{m}$ for the smaller diameter axons and the upper value of 2.5 \textmu $\mathrm{m}$ is more characteristic to the axons with large diameter \cite{Sanders1948}. In peripheral nerves, myelin sheath thickness increases sharply, at first, when the axon radius increases from the smallest physiologically viable diameters and then the thickness increase gradually slows down as the axon diameter approaches the largest physiologically viable values \cite{Sanders1948} in these nerves.

Nodes of Ranvier contain much higher densities of various ion channels than elsewhere on axons. K and Ca channels are about 10~nm {of the radial length across the membrane} and roughly 4~nm {of the longitudinal
diameter in the plane of the membrane} \cite{Doyle1998}. Na channel is about 12~nm {of the radial length across the membrane} and about 10~nm {of the longitudinal
diameter in the plane of the membrane} \cite{Sula2017}. Structurally the myelinated part of the axon is divided into the following regions: next to the node of Ranvier is a region called the paranode. This is the area where the myelin attaches to the axon. The juxtaparanode is located next to the paranode and it is the area where most voltage-gated K$^+$ ion channels are located. The Na$^+$ channels are concentrated in the nodes of Ranvier.
The geometry of a myelinated axon is sketched later in Fig.~\ref{paramS}. {A simplified model of the sheath can be presented as a stack of lipid bilayers.} It should be noted that in the proposed model we do not consider different ion channel distributions between juxtaparanode and paranode and take uniform ion channel distribution within {the} node of Ranvier. 
As a rough ballpark AP propagation velocity for mammals could be taken at around 1~[m/s] (without myelin sheath) \cite{Lodish2004}, however, for example, in the classical Hodgkin-Huxley paper where the HH model was initially introduced \cite{Hodgkin1952} the observed AP propagation velocity is 21.2 [m/s] for the giant axon of the squid. In non-myelinated neurons, the conduction velocity of an action potential is roughly proportional to the diameter of the axon \cite{Lodish2004}. The presence of a myelin sheath around an axon typically increases the velocity of impulse conduction to 10-100~[m/s] \cite{Lodish2004,Schmidt2019}. 
This is a brief overview of the structure of the myelinated axons needed for further modelling. The detailed overview of the properties of myelinated nerve fibres can be found, for example, in \cite{Nave2014,Rosenbluth1999}.

\subsection{Saltatory conduction mechanism hypothesis for myelinated axon}
It is known that the myelination of the axon increases the propagation velocity of the AP and the prevalent explanation in {earlier studies} of this phenomenon is the saltatory conduction mechanism \cite{Bressloff2014,FitzSalt,Tasaki1939,Huxley1949,Frankenhaeuser1964}. 
%{\textbf{NB!} viited 1a,2a,Tasaki 1936, reviewer 1 comment 5.31}
As the capacity of the axon changes significantly between the myelinated and unmyelinated sections, this causes, in a nutshell, the electrical signal to ``jump" between the nodes of Ranvier propagating faster than it would in the case of an unmyelinated axon. The saltatory conduction hypothesis is briefly summarised as follows. %{\color{red}\textbf{TO BE DONE}}

In earlier studies, it is {demonstrated} that myelination increases effective membrane resistance (reduces {the} permeability of ions) and decreases the capacitance of the membrane by several orders of magnitude. As noted, the propagation of AP along myelinated axons is considerably faster than in unmyelinated axons. In the context of the cable equation (which is a starting point for both HH and Lieberstein models) the usual explanation in literature is that transmembrane currents in the myelinated sections can be neglected meaning that the myelin sheath section can be taken in that case as a simple resistor (i.e,~the sheath acts as an insulator). This phenomenon is called a saltatory  (leaping) wave in earlier studies where AP is not propagating continuously along the axon but rather jumps from one node of Ranvier to another node of Ranvier \cite{Bressloff2014}. 

In practice, the saltatory conduction hypothesis, {although} difficult to verify experimentally as measuring sharply localised AP (nodes of Raniver are typically around 1 \textmu $\mathrm{m}$ long) is {measured} \cite{Huxley1949}. What is measured in classical experiments (for example, \cite{Iwasa1980,Tasaki1948,Hodgkin1949}), is AP over some section of the axon (i.e, including the signal from both myelinated {and} non-myelinated sections, moreover most classical experiments are done on non-myelinated axons). However, the proposed hypothesis appears to be a plausible (and {widely used} in earlier studies) starting point to explain why the AP is faster in myelinated axons and we opt to use it as well. It should be noted that in effect we use interpretation which is partly inspired from the continuum mechanics \cite{Sun1968} where we interpret the signal coming from the model not as the evolution of AP in time at a fixed point on axon as is typical but rather a combined signal one might get from a ``unit cell" which contains a node of Ranvier and a myelinated section next to the node, based on assumption that characteristic wavelengths of the signal are much larger than the underlying scale of the micro-structure (nodes of Ranvier in \textmu $\mathrm{m}$ range and myelinated sections typically in hundreds of \textmu $\mathrm{m}$ in length \cite{Lodish2004,Debanne2011}). In comparison, the first harmonics (when looking at the Fourier spectrum of the signal) of the AP in space are from $\mathrm{cm}$ to tens of $\mathrm{cm}$ (depending on the duration and velocity of the signal) \cite{Tamm2021,NovaPeatykk}. 

Following \cite{Bressloff2014}, it is assumed that membrane potential is uniform within a given node of Ranvier (i.e, a node is an isopotential) and denoting the voltage of $n$th node as $V_n$ while treating the adjacent myelinated section as a classical Ohmic resistor with resistance $rL_m$ (where $r$ is intracellular resistance per unit length and $L_m$ is the length of myelinated section) the current $I$ between nodes $n$ and $n+1$ can be written as 
\begin{equation}
I_{n+1}= - \frac{1}{rL_m}\left(V_{n+1}-V_n \right).
\label{salt1}
\end{equation}
Considering the conservation of current at $n$th node implies that the total transmembrane current into that node can be written as
\begin{equation}
2 \pi a l \left( C_m \frac{\partial V_n}{\partial t} + I_{ion} \right) = I_n -I_{n+1} = \frac{1}{r L_m} \left( V_{n+1} - 2 V_{n} + V_{n-1} \right),
\label{salt2}
\end{equation}
where $a$ is the radius of the axon. Bressloff \cite{Bressloff2014} extracts term $\partial V_n / \partial t$ from \eqref{salt2} and constructs equation for the potential $V$ at node $n$ as
\begin{equation}
\frac{\partial V_n}{\partial t} = - \hat{I_{ion}} + D \left( V_{n+1} - 2 V_{n} + V_{n-1} \right),
\label{salt3_0}
\end{equation}
where
\begin{equation}
D = \frac{R_m}{\left( 2 \pi a r \right)l L_m \tau_m } = \frac{\lambda_{m}^{2}}{l L_m \tau_m}.
\label{salt3}
\end{equation}
Here $R=\pi a^2 r$, $\tau_m = R_m C_m$, $\lambda_m = \sqrt{\frac{R_m a}{2 R}}$ and $l$ is the length of the node of Ranvier. Parameter $D$ \eqref{salt3} governs the saltatory conduction velocity between adjacent nodes of Ranvier under the simplifying assumptions done above and is a reasonable starting point. It is worth noting that the philosophy behind constructing eq.~\eqref{salt3_0} is one of the typical approaches how the HH {equations} (which in its ODE form describes signal evolution in time at a fixed spatial point on axon) {are} used to describe travelling AP signal -- meaning that one constructs a pseudo-numerical scheme where adjacent nodes interact and  the signal can propagate that way along the axon. 

We have to emphasize that what follows, is not the direct one-to-one adoption of the saltatory conduction mechanism as proposed in \cite{Bressloff2014} as we are not only considering the electrical effects but also other influences from potentially relevant interactions (like cytoskeleton, for example) and as such, our approach is partly phenomenological. It should be stressed, that we opt to separate myelin geometry along the axon and perpendicular to the axon as theoretically easily observable and propose that the signal propagation velocity from node of Ranvier to node of Ranvier is governed by more than only membrane capacitor dynamics. 
%This way, after the actual propagation velocity is determined from experimental observations, 
{As we know the actual propagation velocity from experimental observations } {(see, for example,} \cite{Arancibia-Carcamo2017}{ and references therein).}
{Knowing AP velocity allows accounting of} the influence of the membrane capacitor dynamics (which can be estimated from the myelination geometry and dielectric properties of the myelin). {Then it is possible to estimate}  other potential influences, like fast enough chemical processes or interactions with the internal structures of the axon to quantify these and hopefully provide a more detailed insight towards functioning and signal propagation in an actual living nerve cell. 

\section{Model for AP propagation on myelinated axon based on Lieberstein model}
\begin{figure}[h]
\includegraphics[width=0.99\textwidth]{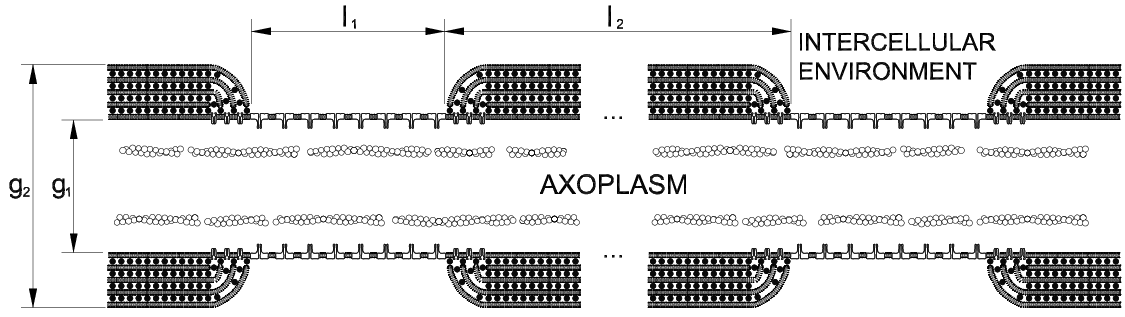}
\caption{Simplified model geometry of the myelinated axon. Here $l_1$ is the length of the node of Ranvier and $l_2$ is the average effective length of the myelinated axon section.}
\label{paramS}
\end{figure}

Taking the elementary form of Maxwell equations combined with the ideas proposed by Lieberstein \cite{Lieberstein1967} in eqs.~\eqref{LIB1} and \eqref{LIB2} as a starting point, we proceed to modify these governing equations to include the effect of myelination on the AP signal propagation. {The simplified geometry of a myelinated axon is shown in} Fig.~\ref{paramS}.
\subsection{Parameters for numerical example}
The following parameters are used: $n=2^{13}$ (number of spatial nodes), $t_{end}=20$ (time in [ms]), $C_m= 1 \, [$\textmu $\mathrm{F / cm^2}]$ (membrane capacitance), $a=0.25 \ldots 32\, [$\textmu $\mathrm{m}]$ (axon radius), $R = 35.4\, [\mathrm{ \Omega \cdot cm}]$ (axoplasm resistance) while the parameters related to ion channel dynamics are the same as taken in \cite{Hodgkin1952}: {$\hat{g}_{Na} = 120 \,[\mathrm{m.mho / cm^2}], \hat{g}_K = 36 \,[\mathrm{m.mho / cm^2}], \hat{g}_l = 0.3 \,[\mathrm{m.mho / cm^2}]$} and $v_{Na} = -115 \,[\mathrm{mV}],\, v_K = 12 \,[\mathrm{mV}],\, v_l = -10.613 \,[\mathrm{mV}]$ while $h_0 = 0.596,\, n_0 = 0.318,\, m_0 = 0.052$ (initial values for parameters $h,\, n,\, m$ in HH {equations} at $t=0$). We take in the following example the $r=R/(\pi a^2)$ (resistance of an axon per unit length). We take $L=22.2 \,[\mathrm{mH \cdot cm}]$ (inductivity) for the numerical example (based on observed AP velocity for the HH model \cite{Hodgkin1952}). Initially we generate a narrow bell-shaped pulse (``spark") for the $v$ in the middle of the space domain which generates the propagating AP. In spatial units, the length of the computation node ($\Delta x$) is 92 $[\mu m]$ while the width of the computational domain in the space (from $n=1$ to $n=8196$) is 24$\pi [\mathrm{cm}] \approx 75.4 [\mathrm{cm}]$. {For the sake of readability, the additional equations and values of physical quantities used in numerical simulation
are collected in Table 1 following} \cite{Hodgkin1952} and \cite{Lieberstein1967}:
%Parameter values collected from} \cite{Hodgkin1952} {and } \cite{Lieberstein1967} { or estimated by authors can be found in Table 1:}
%
%\begin{table}
\\
\begin{center}
\begin{tabular}{ | c | c | c | }
\hline
\multicolumn{3}{|c|}{Table 1: Parameters for finding example solutions for \eqref{LIB1} and \eqref{LIB2}.} \\
\hline
 $\alpha_n = 0.01 \frac{v+10}{\exp(\frac{v+10}{10})-1} $ & $\beta_n = 0.125 \exp(\frac{v}{80}) $ & $\frac{\mathrm{d} n}{\mathrm{d} t} = \alpha_n (1-n) - \beta_n n $ \\ 
\hline
 $\alpha_m = 0.1 \frac{v+25}{\exp(\frac{v+25}{10}) -1} $ & $\beta_m = 4 \exp(\frac{v}{18}) $ & $\frac{\mathrm{d} m}{\mathrm{d} t} = \alpha_m (1-m) - \beta_m m $ \\ 
\hline 
 $\alpha_h = 0.07 \exp(\frac{v}{20}) $ & $\beta_h = \frac{1}{\exp(\frac{v+30}{10})+1} $ & $\frac{\mathrm{d} h}{\mathrm{d} t} = \alpha_h (1-h) - \beta_h h $ \\
\hline
$h_0 = 0.596 $ & $ n_0 = 0.318 $ & $ m_0 = 0.052 $ \\
\hline
$C_a = 0 \left[\frac{\mu \mathrm{F}}{\mathrm{cm}^3}\right]$ & $C_m = 1 \left[\frac{\mu \mathrm{F}}{\mathrm{cm}^2}\right]$ & $R = 35.4 \left[\Omega \cdot \mathrm{cm}\right]$ \\
\hline
$\hat{g}_K = 36 \left[\frac{\mathrm{m.mho}}{\mathrm{cm}^2}\right]$ & $\hat{g}_{Na} = 120 \left[\frac{\mathrm{m.mho}}{\mathrm{cm}^2}\right]$ & $\hat{g}_l = 0.3 \left[\frac{\mathrm{m.mho}}{\mathrm{cm}^2}\right]$\\
\hline
$v_K = 12 \left[\mathrm{mV}\right]$ & $v_{Na} = -115 \left[\mathrm{mV}\right]$ & $v_l=-10.613 \left[\mathrm{mV}\right]$ \\
\hline
%\multicolumn{3}{|c|}{Parameter values from \cite{Hodgkin1952} and  \cite{Lieberstein1967} or estimated by authors.} \\
%\hline
\end{tabular}
\end{center}
%\end{table}
$\quad$\\
It should be stressed, that the chosen parameters do not represent any particular model nerve as some of the parameters are varied (for example, axon radius) in a range that is outside what is typical for the experimentally observed (for the giant axon of the squid) or even taken as a rough estimate (like specific inductivity L). Most of the parameters are taken from \cite{Hodgkin1952} corresponding to giant axon of the squid at about $6.3 [\mathrm{C^\text{o}}]$ while inductivity $L$ is chosen so that the AP signal would have propagation velocity of 21.2 $[\mathrm{m/s}]$ if the axon radius $a$ is 238 [\textmu$m]$.

\begin{figure}[h]
\includegraphics[width=0.99\textwidth]{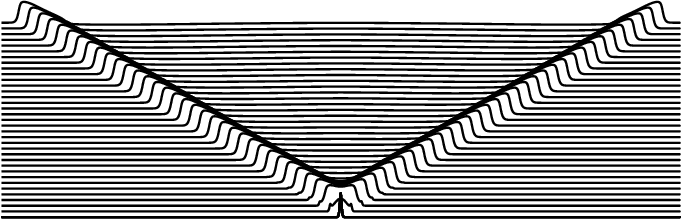}
\caption{Typical solution for Eqs.~\eqref{LIB1} and \eqref{LIB2}. Parameters are the same as in Fig.~\ref{velocity_unmyl}. Horisontal axis is space, vertical axis is time with 0.5 [ms] between the lines.}\label{LIB_timeslice}
\end{figure}

We use pseudospectral method {(PSM)} (see Appendix A in \cite{Raamat2021}) and periodic boundary conditions demonstrating the evolution of solutions for model equations \eqref{LIB1} and \eqref{LIB2}. For initial condition we generate a narrow localised pulse for AP in the middle of the 1D space domain at time $t_0=0$:
\begin{equation}
v(x,t_0) = V_0 \mathrm{sech}^2 (B_0 \cdot x_0), \quad \text{where} \quad x_0 = x - l_0 \cdot \pi,
\end{equation}
where $V_0 = -120$ [mV] is amplitude of the pulse, $B_0$ is the width of the pulse, $l_0$ is number of 2$\pi$ sections in space in space and $x_0$ shifts the hyperbolic secant square function to the middle of the space domain (as it is defined around 0 but we have opted to integrate from 0 to $l_0 \cdot$2$\pi$ for the following numerical example) and take the other quantities initially as zero (at rest). 
Briefly, the main point of the pseudospectral method is that the discrete Fourier transform (DFT) based (PSM) (see also 
\cite{Fornberg1998}) can be used to represent variable $v$  in the Fourier space as
\begin{equation} \label{dft}
\widehat{v}(k,T) = \mathrm{F} \left[ v \right]= \sum^{n-1}_{j=0}{v(j \Delta v, t) \exp{\left(-\frac{2 \pi \mathrm{i} j k}{n} \right)}},
\end{equation}
{where $n$ is the number of space-grid points, $\Delta x=2 \pi/n$ is the space step, $k=0,\pm1,\pm2,\ldots,\pm(n/2-1),-n/2$; $\mathrm{i}$ is the imaginary unit, $\mathrm{F}$ denotes the DFT and $\mathrm{F}^{-1}$ denotes the inverse DFT.
The idea of the PSM is to approximate space derivatives by making use of the DFT}
\begin{equation} \label{dft2}
\frac{\partial^{m} v}{\partial x^{m}} = \mathrm{F}^{-1}\left[(\mathrm{i} k)^{m} \mathrm{F}(v) \right],
\end{equation}
{reducing, therefore, the partial differential equation (PDE) to an ordinary differential equation (ODE) and then to use standard ODE solvers for integration in time. 
For integration in time, the model equations are rewritten as a system of first-order ODE's and a standard numerical integrator is applied. In the numerical examples given in the present paper the ODEPACK FORTRAN code (see }
\cite{ODE}{) ODE solver is used through its NumPy implementation. Handling of the data and initialization of the variables is done in Python by making use of the package SciPy (see }
\cite{SciPy}) and the numerical results are analysed and visualised in the Matlab environment. Example solution can be seen in Fig.~\ref{LIB_timeslice}.

\subsection{Phenomena and descriptions}
We modify the Lieberstein model \cite{Lieberstein1967} to account for the effect of myelination on a nerve fibre, and consider the following {phenomena}:
\begin{enumerate}
\item The velocity of the AP depends on the ratio of lengths between the myelin sheath and the node of Ranvier $\left( l_2/l_1 \right)$ (so-called `$\mu$-ratio' below);
\item The thickness of the myelin sheath affects the velocity of the AP signal (the so-called $g$-ratio) and could be taken into account indirectly through the capacitance variations (included in parameter $\gamma$ below); 
\item The dominant mechanism through which the AP signal velocity in myelinated nerve fibre is increased is the so-called saltatory conduction hypothesis \cite{Bressloff2014};
\item The model equation should be reduced back to the basic model when the myelination approaches to zero (i.e., unmyelinated axon). 
\end{enumerate}
Let us take Lieberstein eqs.~\eqref{LIB1} and \eqref{LIB2}, introducing parameters $\mu$ and $\gamma$ characterizing the AP propagation velocity increase from saltatory conduction \cite{Bressloff2014} and other relevant mechanisms. 
Note that Bressloff \cite{Bressloff2014} has used parameter $D$ that modulates the signal dynamics across the membrane. Here, inspired by Bressloff \cite{Bressloff2014}, we have introduced two parameters: $\gamma$ and $\mu$. The equations {describing the propagation of the AP} are written in the form:
%Note that in the book by Bressloff \cite{Bressloff2014} which we have used as one of the sources of inspiration in regards to saltatory conduction the parameter $D$ \eqref{salt3}, modulates the signal dynamics across the membrane (related to $\gamma$ here). However, the parameter $\mu$ affects the quantity $i_a$ which is the current along the axis of the axon. We can write the governing equations as:
\begin{equation} \label{LIB1m}
\frac{\partial v}{\partial t} + \Phi \cdot \left[ \left( 1 + \gamma \cdot \mu\right) \cdot \frac{\partial i_a}{\partial x} + 2 \pi a \cdot i_m \right] =0,
\end{equation}
\begin{equation} \label{LIB2m}
\frac{\partial i_a}{\partial t} + \frac{\pi a^2}{L} \cdot \left[\frac{\partial v}{\partial x} + r i_a \right]=0,
\end{equation}
\begin{equation} \label{LIB3m}
\Phi=\frac{1}{C_a \pi a^2 + 2 C_m \pi a},
\end{equation}
\begin{equation} \label{LIB4m}
\mu = \frac{l_2}{l_1}, 
\end{equation}
\begin{equation} \label{LIB5m}
i_m =  \hat{g_K} n^4 (v-v_K) + \hat{g_{Na}} m^3 h (v-v_{Na}) + \hat{g_l} (v-v_l).
\end{equation}
\begin{figure}[h]
\includegraphics[width=0.475\textwidth]{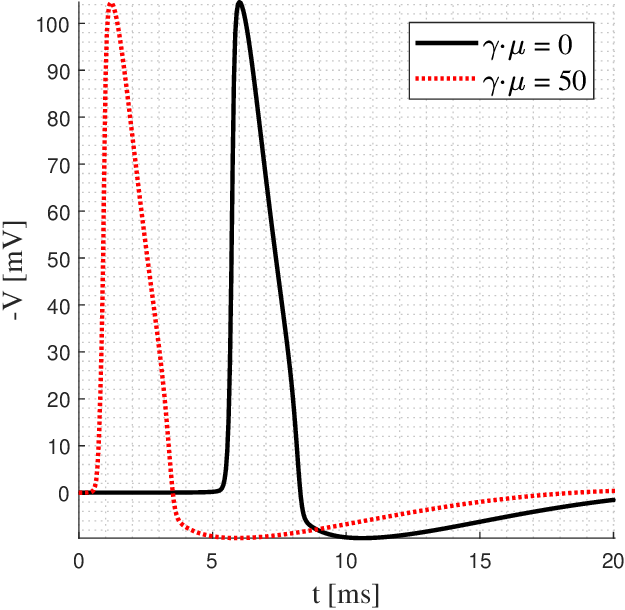}
\includegraphics[width=0.475\textwidth]{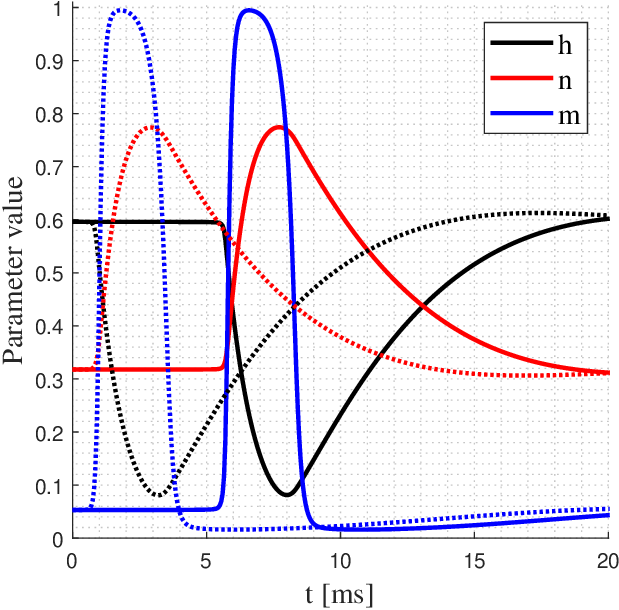}
\caption{AP in time (left panel) and parameters $n,m,h$ in time (right panel) at $n=3996$ (this is $n/2-100$) for $a=1$ [{\textmu}m]. Dotted lines -- myelinated case ($\gamma \cdot \mu = 50$), solid lines -- unmyelinated case ($\gamma \cdot \mu = 0$).}
\label{LibAPintime}
\end{figure}

\begin{figure}[hbt!]
    \includegraphics[width=0.95\textwidth]{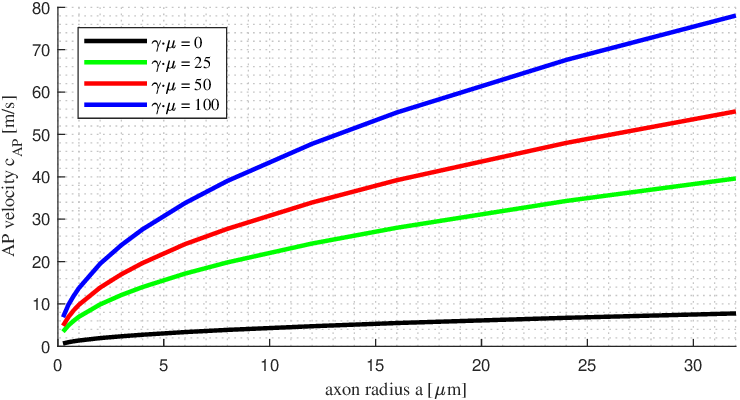}
\caption{Action potential propagation velocity as a function of $\mu$-ratio $\mu=l_2/l_1$ (while $\gamma=1$ here) in the modified Lieberstein model. Model parameters are found in Table 1.}\label{APmylvel}
\end{figure}

In eq.~\eqref{LIB1m} parameter $\mu$ (describing $\mu$-ratio) describes the average length of the myelinated section divided by the average length of the node of Ranvier (see Fig.~\ref{paramS}). It affects the quantity $i_a$ which is the current along the axis of the axon. Parameter $\gamma$ is a phenomenological coefficient which determines conduction velocity between adjacent nodes of Ranvier (generalised from eq.~\eqref{salt3}). Here parameter $\gamma$ includes myelin geometry perpendicular to the axon (related to $g$-ratio). As noted earlier, parameter $\gamma$ is not exactly the same as parameter $D$ (see eq.~\eqref{salt3}) and is a generalised quantity. 
%In our model, this parameter includes the influence of myelin geometry perpendicular to the axon (related to $g$-ratio). 
We assume parameter $\gamma$ to be between 0 and 1: if it is equal to 0, then the current can not propagate between the adjacent nodes (nodes of Ranvier are isolated from each other) and if it is equal to 1, then the myelinated sections are almost perfect conductors 
%(i.e., very high propagation velocity)
{(i.e., the system obeys Ohm's Law)}
 and adjacent nodes react to the changes in a given node almost immediately. 
Term $\gamma \cdot \mu$ could be considered like a generalised dimensionless relative velocity potentially containing all the physical effects (not only resistance but also diffusive, capacitative, inductance effects, the influence of cytoskeleton or even damage or pathological effects, etc) as it is chosen here. One should also stress that in the following example, we take $\gamma=1$ for the sake of simplicity, however, it would be logical that as the length of myelinated sections $l_2$ increases then at some point the parameter $\gamma$ would start to decrease. As a rough initial estimate, following the saltatory conduction hypothesis, one could take $\gamma = 1 - C_{m \mu}/C_{m r}$, where $C_{m \mu}$ is the membrane capacitance of the myelinated section and $C_{m r}$ is the capacitance of the node of Ranvier -- in such a case if we take, for example 1 \textmu$\mathrm{m}$ node of Ranvier, 250 \textmu$\mathrm{m}$ long myelinated section, axon radius of 5 \textmu$\mathrm{m}$ and thickness of the myelin 2.5 \textmu$\mathrm{m}$ the corresponding $\gamma$ would be around 0.5. 
Using similar parameters as in the classical paper by Hodgkin and Huxley \cite{Hodgkin1952} (see Table 1), a numerical example using parameter value $\gamma \cdot \mu=50$ (assuming 1 \textmu$\mathrm{m}$ nodes of Ranvier, 50 \textmu$\mathrm{m}$ length for myelinated sections and taking $\gamma=1$ for the numerical example) and $a=1$ [{\textmu}m] follows {(see Fig.}~\ref{LibAPintime}{).} Choosing $\gamma=1$ (which is an upper limit case) means that we are making here an assumption that saltatory conduction velocity is much faster than signal propagation velocity in an unmyelinated axon. We have to emphasize here that the ``unit cell" we are modelling in the computational node contains both the node of Ranvier and the myelinated section adjacent to it, i.e., the parameters are a mix of ``pure'' Lieberstein model for unmyelinated axon and myelination effects like in \cite{Bressloff2014}. 
%--------------------------------------------------------------------------

One can note that in time (at a fixed spatial node, Fig.~\ref{LibAPintime}) the solutions look practically identical. However, it must be noted that in space the faster (in myelinated axon) AP is shaped significantly wider, as the dynamics of the ion channels are still the same as before, but as the wavefront propagates faster, its dominant wavelength in space is longer. 

The AP propagation velocity as a function of the $\mu$-ratio $\mu= l_2/l_1$ as a function of axon radius (in the $a=0.25 \ldots 32$ [{\textmu}m] range) is depicted in Fig.~\ref{APmylvel}. The case $\mu=0$ corresponds to an unmyelinated axon. It should be noted that here we are assuming that the thickness of the myelin (related to g-ratio, which is included through coefficient $\gamma$ \eqref{salt3}) is the same even if  $\mu$-ratio $\mu$ is varied. The upper limit case ($\gamma=1$) for the conduction velocity is used. It should be noted that the dependence of the AP velocity on the axon radius is still there even in the myelinated case. The larger the radius of the axon, the faster the velocity of the AP increases as the $\mu$-ratio $\mu = l_2/l_1$ is increasing. 

\section{Discussion and summary}

We started with the model for unmyelinated axon \eqref{LIB1}, \eqref{LIB2} \cite{Lieberstein1967,Tamm2025WaMot} in a form which is similar to the Maxwell equations for a transmission line and used the work of Lieberstein as a source of inspiration for handling the question of inductance $L$ in the context of signal propagation along the nerve axon. The key difference between the classical HH model and the Lieberstein-inspired model used here is that the mechanism for signal propagation along the axon emerges {like a wave} as a consequence of opting to keep the inductivity $L$. While, indeed, there exist variations of the classical HH model which support AP signal propagation where the {equations are} written in the form of PDE instead of the usual ODE form (which describes signal evolution in time at a fixed spatial point). In these, normally, the potential-gradient-type member responsible for propagating the signal along the axon is not as clearly defined from {the viewpoint of physics} (usually some kind of abstracted diffusion-type process is used).

We briefly explained the saltatory conduction mechanism hypothesis as we prepared to modify the governing equations to take into account the effect of myelination on the propagation of the AP along the axon. The model \eqref{LIB1}, \eqref{LIB2} based on Maxwell equations for a transmission line is then modified to include the influence of myelination. In addition to the g-ratio normally considered in the earlier studies (taken into account indirectly through coefficient $\gamma$) which takes into account the myelination geometry perpendicular to the axon we introduce the so-called ``myelination-ratio" or $\mu$-ratio which describes the influence of myelin distribution on the signal propagation in the direction of the axis of the axon. The numerical example (see Fig.~\ref{APmylvel}), using parameters from the {earlier studies}, demonstrates physiologically plausible behaviour for the model. The model is reduced to the description of the unmyelinated axon if the length of the myelinated sections along the axon is taken as zero. Under the considered parameter combinations we can observe the AP propagation velocities up to 79 $[\mathrm{m/s}]$ (the signal propagation velocity range for myelinated axons is given as roughly 10 to 100 $[\mathrm{m/s}]$ in the earlier studies \cite{Lodish2004,Schmidt2019}). For comparison, the model yields for the unmyelinated ($\mu=0$) axon in the same parameter range propagation velocity of about 7.8 $[\mathrm{m/s}]$ at $a=32$ [{\textmu}m].

It is important to emphasize that the proposed continuum-based model is philosophically similar to how the transmission line equations are composed. The `unit-cell' in the context of the myelinated axon in the model is composed of the node of Ranvier and the myelinated section next to it. This is opposed to the alternative approach
{where the processes in nodes of Ranvier are described by the classical HH model}
% which is a {standard method taken} in the {studies} 
 while myelinated sections are handled separately either through some numerical scheme or by an alternative model coupled with the HH {equations} in the node of Ranvier through some mechanism. Having a relatively simple pair of PDEs which are connected to the Maxwell equations for anything involving the movement of charges in an environment could be considered superior to investigating causal connections and making predictions than something that is not as clearly connected to the {basic physical considerations}. 
 
\section*{Acknowledgements}

This research was supported by the Estonian Research Council (PRG 1227). J\"uri
Engelbrecht acknowledges the support from the Estonian Academy of Sciences. The authors thank Prof. Antoine Jerusalem from the University of Oxford for helpful comments during the preparation of the present paper. 
%{The authors thank the anonymous reviewers for numerous suggestions for improvements and helpful comments.}

\section*{Author contributions}
Kert Tamm contributed by writing manuscript text and composing figures, performing numerical simulations and analysis of the results. Tanel Peets contributed analysis for comparing the modified Lieberstein equation against the classical Hodgkin-Huxley {equations}. All other work (like novel ideas presented, physical interpretation of the results, proofreading, etc) is a collaborative effort equally contributed by all authors. 

%\appendix
%\section{Example Appendix Section}
%\label{app1}

%%       \cite{<label>} ==> [1]

%%
%Example citation, See \cite{lamport94}.




%\begin{thebibliography}{00}
%
%%% For numbered reference style
%%% Text of bibliographic item
%
%\bibitem{lamport94}
%  \textit{\LaTeX: a document preparation system},
%  Addison Wesley, Massachusetts,

%\end{thebibliography}

\bibliographystyle{abbrv}
\bibliography{library}

\end{document}